\documentclass{aa}  

\usepackage{graphicx}
\usepackage{txfonts}
\usepackage{units}
\usepackage{soul}
\setstcolor{red}
\usepackage{keyval}
\usepackage{comment}
\usepackage{colortbl}
\usepackage{natbib}
\usepackage{hyperref} 
\usepackage{xcolor}
\usepackage{xspace}
\usepackage{pdflscape}
\usepackage{placeins}

\newcommand{\rahms}[4]{$#1^{\rm h}#2^{\rm m}#3\mbox{$^{\rm s}\mskip-7.6mu.\,$}#4$}
\newcommand{\decdms}[4]{$#1^{\circ}#2'#3\mbox{$''\mskip-7.6mu.\,$}#4$} 

\newcommand{\msun}{~M\mbox{$_\odot$}\xspace}
\newcommand{\gaia}{\textit{Gaia}\xspace}
\newcommand{\mpy}{{mas\,yr$^{-1}$}\xspace}

\newcommand{\acca}{$\dot{\mu}_\alpha^*$\xspace}
\newcommand{\accd}{$\dot{\mu}_\delta$\xspace}

\usepackage{tikz}
\definecolor{lime}{HTML}{A6CE39}
\DeclareRobustCommand{\orcidicon}{%
    \begin{tikzpicture}
    \draw[lime, fill=lime] (0,0) 
    circle [radius=0.16] 
    node[white] {{\fontfamily{qag}\selectfont \tiny ID}};
    \draw[white, fill=white] (-0.0625,0.095) 
    circle [radius=0.007];
    \end{tikzpicture}
    \hspace{-2mm}
}

\newcommand{\orcidSAD}{\href{https://orcid.org/0000-0001-6010-6200}{\orcidicon}}
\newcommand{\orcidMK}{\href{https://orcid.org/0000-0002-5365-1267}{\orcidicon}}
\newcommand{\orcidJO}{\href{https://orcid.org/0000-0001-7776-498X}{\orcidicon}}
\newcommand{\orcidLL}{\href{https://orcid.org/0000-0002-5635-3345}{\orcidicon}}
\newcommand{\orcidGO}{\href{https://orcid.org/0000-0002-2863-676X}{\orcidicon}}
\newcommand{\orcidPG}{\href{https://orcid.org/0000-0003-2271-9297}{\orcidicon}}
\newcommand{\orcidJM}{\href{https://orcid.org/0000-0002-1963-6848}{\orcidicon}}
\newcommand{\orcidLF}{\href{https://orcid.org/0000-0003-2737-5681}{\orcidicon}}
\newcommand{\orcidAM}{\href{https://orcid.org/0000-0002-2564-3104}{\orcidicon}}
\newcommand{\orcidJF}{\href{https://orcid.org/0000-0001-8694-4966}{\orcidicon}}

\defcitealias{kounkel2017}{KHL17}
\defcitealias{dzib2025}{Paper~II}

\begin{document}

\title{Dynamical masses of YSOs with the VLBA: DYNAMO VLBA}
\subtitle{Trigonometric parallaxes and proper motions of YSOs in Orion}

\titlerunning{DYNAMO-VLBA: Astrometry to YSOs in Orion}

\author{
  Sergio A. Dzib\orcidSAD\inst{1} \and Jazm\'{\i}n Ord\'{o}\~nez-Toro\orcidJO\inst{2,3} \and Laurent Loinard\orcidLL\inst{2,4,5} \and 
  Marina Kounkel\orcidMK\inst{6} \and \\ Gisela Ortiz-Leon\orcidGO\inst{7} \and Phillip A. B. Galli\orcidPG\inst{8} \and 
  Luis F. Rodr\'{\i}guez\orcidLF\inst{2} \and Amy J. Mioduszewski\orcidAM\inst{9} \and \\ Josep M. Masqu\'e\orcidJM\inst{10,11}  \and Eoin O'Kelly\inst{12} \and Jan Forbrich\orcidJF\inst{12,13} \and Karla Moo-Herrera\inst{14}
}

\institute{
  \inst{1}{ Max-Planck-Institut f\"ur Radioastronomie, Auf dem H\"ugel 69, D-53121 Bonn, Germany;} \email{sdzib@mpifr-bonn.mpg.de}\\
  \inst{2}{ Instituto de Radioastronom\'{\i}a y Astrof\'{\i}sica, Universidad Nacional Aut\'onoma de M\'exico, Morelia 58089, M\'exico}\\
  \inst{3}{ Departamento de Astronom\'{\i}a, Universidad de Guanajuato, Apartado Postal 144, 36000 Guanajuato, M\'exico}\\
  \inst{4}{ Black Hole Initiative at Harvard University, 20 Garden Street, Cambridge, MA 02138, USA}\\
  \inst{5}{ David Rockefeller Center for Latin American Studies, Harvard University, 1730 Cambridge Street, Cambridge, MA 02138, USA}\\
  \inst{6}{ Department of Physics and Astronomy, University of North Florida, 1 UNF Dr., Jacksonville, FL, 32224}\\
  \inst{7}{ Instituto Nacional de Astrofísica, Óptica y Electrónica, Apartado Postal 51 y 216, 72000 Puebla, México}\\
 \inst{8}{ Instituto de Astronomia, Geofísica e Ciências Atmosféricas, Universidade de São Paulo, Rua do Matão, 1226, Cidade Universitária, 05508-090, São Paulo-SP, Brazil.}\\
 \inst{9}{ National Radio Astronomy Observatory, Domenici Science Operations Center, 1003 Lopezville Road, Socorro, NM 87801, USA}\\
 \inst{10}{ Departament de Física Quàntica i Astrofísica (FQA), Universitat de Barcelona (UB), c/ Martí i Franquès 1, 08028 Barcelona, Catalunya, Spain}\\
 \inst{11}{ Institut de Ciències del Cosmos (ICCUB), Universitat de Barcelona (UB), c/ Martí i Franquès 1, 08028 Barcelona, Catalunya, Spain}\\
 \inst{12}{ Centre for Astrophysics Research, University of Hertfordshire, College Lane, Hatfield, AL10 9AB, UK}\\
 \inst{13}{ Center for Astrophysics | Harvard \& Smithsonian, 60 Garden St, MS 72, Cambridge, MA 02138, USA}\\
 \inst{14}{ Facultad de Ingeniería, Universidad Aut\'onoma de Yucat\'an. Avenida Industrias No Contaminantes por Anillo Periférico Norte s/n, 97302, Mérida, Yucatán, México}\\
  }

\date{Received ; accepted }

\abstract
{We present results from a multi-epoch Very Long Baseline Array (VLBA) survey of compact radio sources in the Orion complex, conducted within both the DYNAMO–VLBA and the GOBELINS projects.  
Our observations detected 216 compact radio sources, of which 58 yielded reliable multi-epoch astrometric solutions.  
For these sources, we derived trigonometric parallaxes and proper motions with typical precisions of about 0.05~mas and 0.10~mas yr$^{-1}$, respectively.  
The measured parallaxes range between 2.26 and 2.65 mas, corresponding to distances of 380–440 pc, and delineate the depth of the Orion star-forming complex.  
We determine mean distances of $405\pm16$ pc for NGC 2068, $403\pm5$ pc for NGC 2024, $407\pm12$ pc for the $\sigma$ Orionis region, $388.5\pm1.7$\,pc for the Orion Nebula Cluster (ONC), and $438\pm12$~pc for L1641. 
A comparison with \textit{Gaia} DR3 astrometry for 28 common sources reveals negligible mean parallax offsets ($\Delta\varpi=-0.02\pm0.01$\,mas) and small systematic differences in proper motions ($\sim$0.07\,\mpy), likely due to residual rotation of the \textit{Gaia} reference frame.  
Our results demonstrate the capability of high-precision radio astrometry to map embedded stellar populations and to provide an independent calibration of the \textit{Gaia} reference system in obscured regions.
}

\keywords{astrometry --- stars:formation --- stars:kinematics
}

\maketitle
\section{Introduction}
Young stellar objects (YSOs) are often deeply embedded within their natal molecular clouds, where high extinction hinders optical astrometry.  
Radio interferometry offers a complementary means to locate such sources and to trace their kinematics with sub-milliarcsecond precision.  
Gyrosynchrotron emission from magnetically active coronae produces compact, non-thermal radio sources that are ideally suited for Very Long Baseline Interferometry (VLBI) observations, allowing trigonometric parallaxes and proper motions to be measured with micro-arcsecond accuracy \citep[e.g.,][]{loinard2007,Dzib2011}.

Over the past decade, advances in wide-field VLBI correlation, particularly the implementation of the \textsc{DiFX} software correlator \citep{Deller2011}, have enabled large multi-epoch surveys targeting hundreds of radio-emitting YSOs within a single primary beam.  
These surveys, such as GOBELINS \citep{loinard2011,ortiz2017a}, have provided high-precision astrometry for nearby star-forming regions, complementing and extending the \gaia mission to heavily obscured environments.  

The Orion molecular complex represents one of the richest laboratories for studying young stellar populations across a wide range of masses and evolutionary stages.
Its distance structure has been investigated through both optical and radio VLBI observations \citep[e.g.,][]{kounkel2017,grossscheld2018}.
The {Dynamical Masses of multiple YSOs with the VLBA} (DYNAMO–VLBA) project builds upon these previous efforts by obtaining new VLBA observations aimed primarily at monitoring the orbital motions of binary YSOs in nearby star-forming regions.
In addition, these observations provide the opportunity to measure precise parallaxes and proper motions not only for known binaries, but also for all compact radio sources within each observed field.

In this first paper of the Orion DYNAMO–VLBA series, we focus on the astrometric results for the full sample of detected radio sources, independent of their multiplicity.  
We combine our new observations with archival data from the Gould's Belt Distances Survey (GOBELINS) project to derive updated distances to the main Orion subregions and to compare our results with \gaia\,DR3 astrometry.  A companion paper \citep[][hereafter \citetalias{dzib2025}]{dzib2025} will present the detailed orbital analysis and dynamical masses of the binary and multiple systems.

\begin{table*}
\small
\begin{center}
\renewcommand{\arraystretch}{1.0}
\caption{Observed blocks.}
\begin{tabular}{c|cccccccccccccc}\hline\hline
BD215& \multicolumn{2}{c}{Position}& Correlated &Cadence&MPC & SPC1 & SPC2 &SPC3\\
Block&       R.A.&Dec.             & positions& (days) &\\ 
\hline
F &  \rahms{05}{35}{20}{16} &\decdms{-05}{20}{06}{0}& 87&  60&J0539--0514&J0529--0519&J0541--0541&J0532--0307\\
G &  \rahms{05}{35}{31}{37} &\decdms{-05}{16}{02}{6}& 3 &  60&J0539--0514&J0529--0519&J0541--0541&J0532--0307\\
H &  \rahms{05}{35}{21}{32} &\decdms{-05}{12}{12}{7}& 4 & 180&J0539--0514&J0529--0519&J0541--0541&J0532--0307\\
I &  \rahms{05}{39}{36}{54} &\decdms{-02}{42}{17}{2}& 1 &  60&J0532--0307&J0539--0514&J0529--0519&J0558--0055\\
J &  \rahms{05}{41}{37}{74} &\decdms{-01}{53}{51}{6}& 6 &  60&J0532--0307&J0539--0514&J0529--0519&J0558--0055\\
K &  \rahms{05}{41}{46}{16} &\decdms{-01}{56}{22}{2}& 6 & 180&J0532--0307&J0539--0514&J0529--0519&J0558--0055\\
L &  \rahms{05}{46}{43}{39} &\decdms{+00}{04}{36}{0}& 3 & 120&J0558--0055&J0600--0005& J0552+0313&J0532--0307\\
\hline\hline
\end{tabular}\label{tab:blocks}
\end{center}
\tablefoot{Columns are (left to right): The name of the block, R.A. and Dec. of the center of the field, number of targets (i.e. correlated position in the corresponding field), a mean time of the separation between observed epochs, and main and secondary phase calibrators of the corresponding field. }
\end{table*}

\section{Data and analysis methods}\label{sec:data}

\subsection{DYNAMO-VLBA observations}

The DYNAMO-VLBA observations were obtained under the VLBA project code BD215.
For sources in the Orion constellation, we divided our target sources in seven 
distinct blocks (named from F to L, see~Table~\ref{tab:blocks}). These blocks were chosen 
to observe known and suspected binaries that were reported by \citep[][hereafter \citetalias{kounkel2017}]{kounkel2017}, 
and are known radio sources. Additionally, each block targeted 
radio sources located inside the primary beam of the antennas, which mostly consisted of 
other known YSOs and other radio sources of unknown origin. The blocks were observed 
a total of six times (labeled from 0 to 5), and the cadence of the observation per block was 
chosen according to the suspected period of known binaries in the field from GOBELINS 
results. In total, there are 42 DYNAMO-VLBA observations of targets in Orion, whose 
epochs are listed in Table~\ref{tab:Depochs}.

The observations were obtained with the VLBA telescope at a central frequency of 5.0 GHz ($\lambda$ = 6.0~cm) with a total bandwidth of 256 MHz.
The total duration of each observation was three hours. The first 25~minutes were spent observing 
around a dozen quasars distributed over the entire sky (i.e., a geodetic block). 
The observation continues by pointing three secondary phase calibrators (SPCs), which
are observed with one-minute scans. Then, a two-minute scan on the target source
bracketed with one-minute scans on a nearby main phase calibrator (MPC) are observed. This cycle is repeated for 
30 minutes, when the SPCs are observed once more.
The cycle SPCs[3 min] -- (MPC[1 min]-target[2 min])[30 min] -- SPCs[3 min] is repeated a few times. 
The last 25 minutes are used to observe a second geodetic block, completing the three-hour observation. 
All MPCs and SPCs used in these observations are listed in Table~\ref{tab:PhCal}, together with the position assumed for correlation. These positions are consistent with those used by \citetalias{kounkel2017} within the GOBELINS project. 

The data calibration followed a standard scheme for astrometric phase-referenced VLBI observations and included multiple secondary phase calibrators and geodetic group-delay blocks. Residual atmospheric and ionospheric phase gradients were mitigated using the AIPS task ATMCA \citep{atmca}, which exploits the information from multiple calibrators to improve phase transfer to the target sources.

This calibration strategy follows the core principles of multi-calibrator (or MultiView) phase referencing, in which spatially distributed calibrators are used to reduce residual atmospheric and ionospheric systematics, particularly important at frequencies around 5 GHz \citep[e.g.,][]{loinard2007,dzib2010,ortiz2017a}. While recent dedicated MultiView implementations explicitly model ionospheric phase wedges to further improve astrometric accuracy \citep[e.g.,][]{rioja2017,hyland2022}, we do not perform such post-correlation corrections here. Instead, the use of multiple secondary calibrators together with ATMCA provides redundancy in phase transfer and significantly reduces systematic errors, as supported by the empirical error analysis presented in Sect. 3.4.

\begin{table}
\small
\begin{center}
\renewcommand{\arraystretch}{1.01}
\caption{DYNAMO VLBA observed epochs.}
\begin{tabular}{c|cccccccccccccc}\hline\hline
BD215& Obs. date               &  Julian &Noise \\
Epoch& {\footnotesize (DD/MM/YYYY hh:mm)} &Date     &($\mu$Jy bm$^{-1})$ \\ 
\hline
F0&17/02/2018 01:55&2458166.58&47\\
F1&28/04/2018 21:16&2458237.39&58\\
F2&01/07/2018 17:03&2458301.21&48\\
F3&03/09/2018 12:53&2458365.04&54\\
F4&07/11/2018 08:37&2458429.86&45\\
F5&06/01/2019 04:41&2458489.70&49\\
G0&22/02/2018 01:35&2458171.57&32\\
G1&01/05/2018 21:04&2458240.38&28\\
G2&04/07/2018 16:53&2458304.20&34\\
G3&10/09/2018 12:25&2458372.02&42\\
G4&10/11/2018 08:25&2458432.85&27\\
G5&07/01/2019 04:37&2458490.69&30\\
H0&04/03/2018 12:52&2458182.04&27\\
H1&07/09/2018 12:37&2458369.03&28\\
H2&02/03/2019 13:01&2458545.04&25\\
H3&01/09/2019 13:01&2458728.04&26\\
H4&18/03/2020 11:52&2458926.99&34\\
H5&06/10/2020 10:41&2459128.95&28\\
I0&09/03/2018 12:33&2458187.02&24\\
I1&08/05/2018 20:37&2458247.36&17\\
I2&07/07/2018 16:42&2458307.20&20\\
I3&08/09/2018 12:34&2458370.02&25\\
I4&09/11/2018 08:31&2458431.85&25\\
I5&09/01/2019 04:31&2458492.69&22\\
J0&11/03/2018 12:23&2458189.02&38\\
J1&20/05/2018 19:49&2458259.33&26\\
J2&30/07/2018 15:11&2458330.13&33\\
J3&24/09/2018 11:30&2458385.98&53\\
J4&16/11/2018 08:02&2458438.83&34\\
J5&12/01/2019 04:18&2458495.68&27\\
K0&01/04/2018 11:05&2458209.96&24\\
K1&18/10/2018 09:56&2458409.91&27\\
K2&23/04/2019 21:36&2458597.40&42\\
K3&18/10/2019 09:59&2458774.92&27\\
K4&04/04/2020 22:48&2458944.45&34\\
K5&13/10/2020 10:14&2459135.93&32\\
L0&07/04/2018 22:37&2458216.44&31\\
L1&18/08/2018 13:55&2458349.08&41\\
L2&11/12/2018 06:23&2458463.77&52\\
L3&22/04/2019 21:40&2458596.40&43\\
L4&01/08/2019 15:02&2458697.13&34\\
L5&09/12/2019 06:32&2458826.77&42\\
\hline\hline
\end{tabular}\label{tab:Depochs}
\end{center}
\tablefoot{Columns are (left to right): Code of epoch, which is related to its block, date and time of the middle of the observation, corresponding Julian date, and the mean noise of that epoch. }
\end{table}

\begin{table}
\small
\begin{center}
\setlength{\tabcolsep}{3pt}
\renewcommand{\arraystretch}{1.0}
\caption{Calibrators and the pointed positions in our observations.}
\begin{tabular}{c|cccccccccccccc}\hline\hline
Name&  R.A.&Dec.&$\Delta\alpha_{RFC}$&$\Delta\delta_{RFC}$\\ 
\hline
J0539--0514& \rahms{05}{39}{59}{937192} &\decdms{-05}{14}{41}{30174}&--0.54&1.01\\
J0529--0519& \rahms{05}{29}{53}{533450} &\decdms{-05}{19}{41}{61678}&0.81&--0.58\\
J0541--0541& \rahms{05}{41}{38}{083371} &\decdms{-05}{41}{49}{42843}&0.04&--0.19\\
J0532--0307& \rahms{05}{32}{07}{519261} &\decdms{-03}{07}{07}{03799}&0.97&0.93\\
J0558--0055& \rahms{05}{58}{44}{391460} &\decdms{-00}{55}{06}{92375}&1.92&--4.78 \\
J0600--0005& \rahms{06}{00}{03}{503368} &\decdms{-00}{05}{59}{03477}&0.33&0.46 \\
J0552+0313&  \rahms{05}{52}{50}{101499} &\decdms{+03}{13}{27}{24311}&--0.76&1.10 \\
J0542--0913& \rahms{05}{42}{55}{877408} &\decdms{-09}{13}{31}{00660}&1.04&0.69 \\
\hline\hline
\end{tabular}\label{tab:PhCal}
\end{center}
\tablefoot{Columns are (left to right): The calibrator name, the celestial position, the position shift (in mas) between the correlated position and the RFC 2025B position.}
\end{table}

After calibration, the data were imaged using a pixel size
of 200$\,\mu$as and a square image of size 4096 pixels (i.e.,
$0\rlap{.}''89\times0\rlap{.}''89$). Then, we proceeded to 
image the detected brightest emission peak by centering the new image
on it, with a pixel size of 100$\,\mu$as and a square image of size 
1024 pixels (i.e., $0\rlap{.}''1\times0\rlap{.}''1$). Both images 
used a natural weighting scheme (Robust=5 in AIPS). 
The mean noise level in the final images is in the range 
20-50$\,\mu$Jy beam$^{-1}$ (see Table~\ref{tab:Depochs}). 

The positions and flux densities of the detected sources were measured using a two-dimensional Gaussian ellipsoid fitting procedure (AIPS task JMFIT). The measured flux densities are corrected for primary beam attenuation using the same methodology as in other multi-phase-center VLBA experiments \citep[e.g.,][ O'Kelly et al. in prep.]{Deane2024}. The primary-beam response was approximated by an Airy pattern corresponding to a uniformly illuminated circular antenna,

\begin{equation}\label{eqAiry}
P(\theta)=\left( \frac{2J_1(x)}{x}\right)^{2};\qquad x=\frac{\pi D{\theta}}{\lambda}.
\end{equation}

\noindent In this equation, $J_1(x)$ is the Bessel function of order one, $D$ is the antenna diameter, $\lambda$ is the observing wavelength, and $\theta$ is the source angular offset from the pointing center. For the VLBA, we adopted an effective antenna diameter of $D=25.48$\,m \citep{Middelberg2013}. The corrected flux densities were then obtained by dividing the measured values by $P(\theta)$.

Finally, it is worth noticing 
at this point that source positions have different contributing errors.
First, the statistical errors from the fitting. Second, systematic errors
as residuals from the phase transfer from the MPC to the target source. 
These systematic errors are estimated to be of the order of 
$0.1$\,mas per each degree of separation between them. 
Finally, radio AGN and quasar positions are constantly being 
refined with additional VLBI observations from programs of 
geodesy and absolute astrometry and constitute the Radio Fundamental Catalog 
(RFC) whose positions are tied to the International Celestial Reference Frame 
(ICRF); a detailed description is given by \citet{petrov2025}. The positions
of the MPCs (see Table~\ref{tab:PhCal}) used in our observations were obtained 
more than a decade ago, and differ from their more accurate values known to
date (listed in Table~\ref{tab:NPhCal}). This difference is reflected in the
final position of the targets which are expected to be shifted by the same positional differences of the MPC. In Table~\ref{tab:PhCal}, we also list the difference between the correlated position, which we will use to correct the
position of the detected radio sources after astrometric fitting.

\subsection{GOBELINS data}

To complement our astrometric analysis, we have used VLBA observations reported 
by \citetalias{kounkel2017}, obtained as part of the GOBELINS project 
\citep[][Loinard et al. in prep.]{ortiz2017a}. We note that the GOBELINS project 
observed a total of 46 fields targeting 300 radio sources in Orion previously uncovered with 
the VLA by \citep{kounkel2014}.
Source properties were reported by \citetalias{kounkel2017}, and we have used the 
source positions listed in their Table~4. Additional, as-yet unpublished, observations 
of Orion sources were obtained as part of the GOBELINS project after the 
publication of \citetalias{kounkel2017}. These additional GOBELINS observations were calibrated 
and imaged using the same procedure described above. Details on these observed epochs
are listed in Table~\ref{tab:Gepochs}.

DYNAMO-VLBA and GOBELINS observations used the same observation strategy, where the 
common stars in both projects were observed with the same MPC and SPCs. 
The MPC and SPC positions for correlation were also the same in both
projects, also see Table~2 in \citepalias{kounkel2017}. This simplifies the astrometric 
combination of the two data sets. It is important to note, however, that the flux densities reported by \citetalias{kounkel2017} were not corrected for primary beam attenuation, whereas such corrections were applied in the present work.

\begin{table}
\small
\begin{center}
\renewcommand{\arraystretch}{1.0}
\caption{Previously unpublished GOBELINS observations.}
\begin{tabular}{c|cccccccccccccc}\hline\hline
     & Obs. date               &  Julian &Noise \\
Epoch& {\footnotesize (DD/MM/YYYY hh:mm)} &Date     &($\mu$Jy bm$^{-1})$ \\ 
\hline
I0&16/03/2016 01:13&2457463.55&33\\
IQ&12/10/2016 11:24&2457673.98&32\\
IR&15/10/2016 11:35&2457676.98&34\\
IS&16/10/2016 11:09&2457677.96&34\\
IT&17/10/2016 11:05&2457678.96&60\\
J7&16/01/2017 05:29&2457769.73&30\\
J8&18/01/2017 04:59&2457771.71&26\\
J9&24/01/2017 04:36&2457777.69&69\\
JI&25/03/2017 01:07&2457837.55&25\\
JJ&08/04/2017 00:10&2457851.51&27\\
JK&10/04/2017 23:58&2457854.5&56\\
K0&06/07/2017 18:15&2457941.26&47\\
K1&13/07/2017 17:51&2457948.24&32\\
K2&14/07/2017 17:45&2457949.24&29\\
KF&22/08/2017 15:27&2457988.14&30\\
KD&27/08/2017 14:51&2457993.12&44\\
KE&01/09/2017 14:35&2457998.11&28\\
\hline\hline
\end{tabular}\label{tab:Gepochs}
\end{center}
\tablefoot{Columns are the same as in Table \ref{tab:Depochs}. }
\end{table}

\subsection{Astrometric fitting equations and procedures}\label{subs:fits}

The fitting of the motion of the detected sources was performed according to the complexity of the system. We have classified the sources into four main categories: (1) single stars, (2) astrometric binaries, (3) visual radio binaries, and (4) known spectroscopic or visual binaries.

For single sources, the trajectory on the plane of the sky has been fitted using a combination of proper motion and trigonometric parallax. The equations used in this case are:
\begin{align}
\alpha(t) &= \alpha_0 + \mu_\alpha \cdot \cos(\delta) \, t + \varpi \cdot f_\alpha(t), \\
\delta(t) &= \delta_0 + \mu_\delta \cdot t + \varpi \cdot f_\delta(t),
\end{align}

\noindent
where $(\alpha_0, \delta_0)$ are the reference coordinates at a given epoch, $(\mu_\alpha, \mu_\delta)$ are the components of the proper motion, $\pi$ is the parallax, and $(f_\alpha, f_\delta)$ represents the parallactic factors projected along right ascension and declination. These projections depend on the barycentric position of Earth and the direction to the source \citep{Seidelmann1992}.

For astrometric binaries, where only one radio source is detected but which exhibit a non-linear trajectory revealing the presence of an unseen companion, an acceleration term has been added to the model. This approach is appropriate for systems with orbital periods significantly longer than the time span of our observations (typically $<$ 5 years). In such cases, it is reasonable to assume a constant acceleration \citep[see, e.g.,][]{loinard2007}, and the motion is described by:

\begin{align}
\alpha(t) &= \alpha_0 + \mu_\alpha \cdot \cos(\delta) \, t + \frac{1}{2} \dot{\mu}_\alpha\cdot \cos(\delta) \cdot t^2 + \varpi \cdot f_\alpha(t), \\
\delta(t) &= \delta_0 + \mu_\delta \cdot t + \frac{1}{2} \dot{\mu}_\delta \cdot t^2 + \varpi \cdot f_\delta(t),
\end{align}

\noindent
where  $\dot{\mu}_\alpha\cdot \cos(\delta)$ (=\acca) and \accd are the components of the acceleration in right ascension and declination, respectively.

In the case of visual radio binaries, where both components are detected simultaneously in radio images at multiple epochs, the position of each component has been fitted by combining the astrometric parameters with the orbital motion around the center of mass of the system:

\begin{align}
\alpha(t) &= \alpha_0 + \mu_\alpha \cdot \cos(\delta) \, t + \varpi \cdot f_\alpha(t) + a_1 \cdot Q_\alpha(t), \\
\delta(t) &= \delta_0 + \mu_\delta \cdot t + \varpi \cdot f_\delta(t) + a_1 \cdot Q_\delta(t),
\end{align}

\noindent
where $Q_\alpha(t)$ and $Q_\delta(t)$ are the projections of the orbital motion along right ascension and declination, respectively, and $a_1$ is the projected semimajor axis of the orbit for the primary component. When the secondary component is also detected, the same model is applied using $a_2$, taking into account that both components trace the same orbit but are $180^\circ$ out of phase with respect to the barycenter.

For known binaries detected as single radio sources, we can use information from the literature (e.g., from spectroscopic observations) to fit the orbital motion of the detected component around the center of mass of the system. In most cases, spectroscopic binaries provide the mass ratio $q = M_2/M_1$, but they can also provide other astrometric parameters that can be fixed in our astrometric fits. This approach allows us to estimate the individual stellar masses, even when only one component is detected.

\begin{figure*}[!th]
    \centering
    \includegraphics[width=0.95\textwidth]{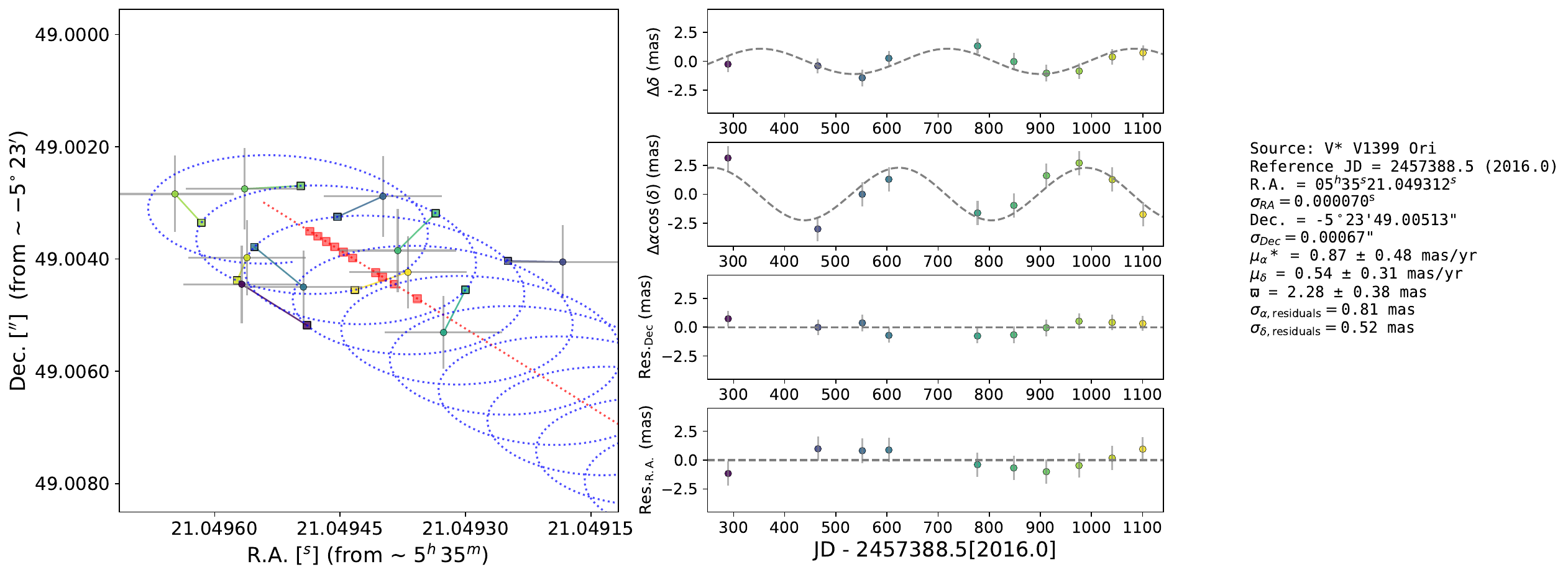}
    \caption{Astrometric motion of V*~V1399~Ori. {Left:} Filled circles indicate the measured positions, with $1\sigma$ uncertainties.  The blue-dotted curve shows the best-fit linear model to the motion, with filled squares indicating the model positions at the observed epochs. Measured and model positions are color-coded by observing epoch, and line segments connect the corresponding epochs.
    The red dotted line represents the motion after subtracting the trigonometric parallax, with red squares indicating the corresponding model positions. {Central panels:} position offsets in declination (top) and right ascension (second panel) as a function of time, obtained after subtracting the best-fit linear proper motion, together with the best-fit parallax model (dashed lines). The two lower panels show the post-fit residuals in declination and right ascension. Symbols are color-coded by observing epoch, as in the left panel. Right: Parameters derived from the linear motion fit.  } 
    \label{fig:NL}
\end{figure*}
\begin{figure*}[!h]
    \centering
\includegraphics[width=0.9\linewidth]{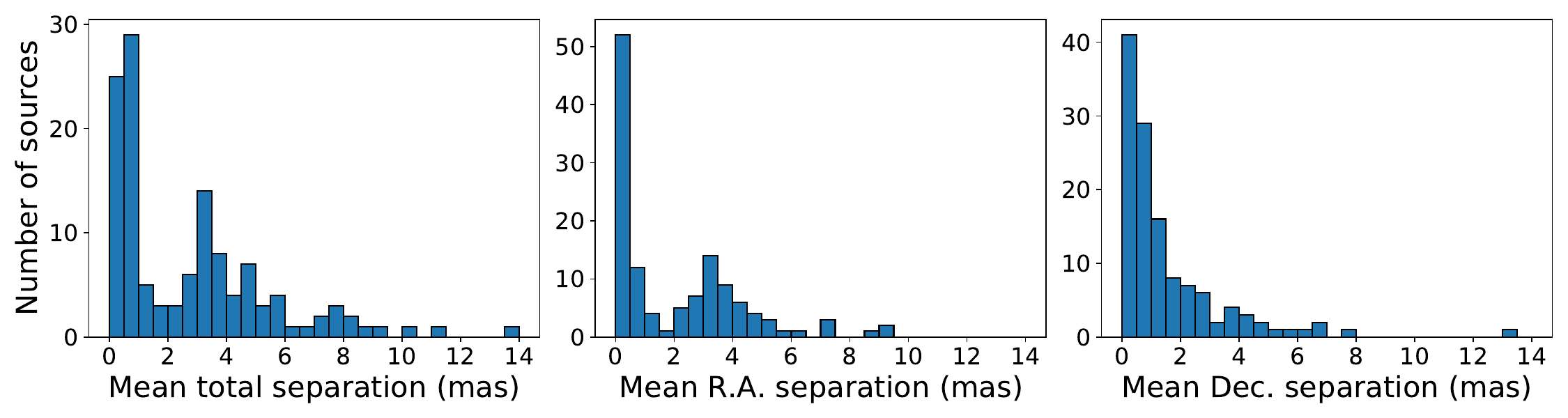}
    \caption{Histograms showing the separation of sources through epochs.}
    \label{fig:Sep}
\end{figure*}

\begin{table*}[!ht]
\begin{center}
\small
\footnotesize
\scriptsize
\setlength{\tabcolsep}{5pt}
\renewcommand{\arraystretch}{1.0}
\caption{Properties of detected radio sources.}\label{tab:full}
\begin{tabular}{ccccccccccccccccccc}
\hline\hline
     &   VLBA &          Julian &     R.A. &      $\sigma_\alpha$ &  Dec. & $\sigma_\delta$ &    Flux  &\\
Name &   Project &     Date      &   ($^{h}\,^{m}\,^{s}$) &  ($^{s}$)    & ($^\circ\,'\,''$)&  ($''$) &   ($\mu$Jy)& MPC\\
\hline
GBS-VLA J053428.42+100422.6 & BL175GO & 2456733.58 & $05\, 34\, 28.418538$ & $0.000034$ & $+10\, 04\, 22.54831$ & $0.00075$ & $1838 \pm 271$ & J0536+0944 \\
GBS-VLA J053428.42+100422.6 & BL175FX & 2457294.01 & $05\, 34\, 28.418489$ & $0.000010$ & $+10\, 04\, 22.55126$ & $0.00025$ & $1421 \pm 202$ & J0536+0944 \\
GBS-VLA J053455.56+095713.2 & BL175GO & 2456733.58 & $05\, 34\, 55.558763$ & $0.000060$ & $+09\, 57\, 13.11874$ & $0.00126$ & $1463 \pm 387$ & J0536+0944 \\
GBS-VLA J053503.54+095640.8 & BL175GO & 2456733.58 & $05\, 35\, 03.536585$ & $0.000045$ & $+09\, 56\, 40.76773$ & $0.00100$ & $970 \pm 220$ & J0536+0944 \\
GBS-VLA J053503.54+095640.8 & BL175ER & 2456950.00 & $05\, 35\, 03.536520$ & $0.000023$ & $+09\, 56\, 40.77070$ & $0.00125$ & $543 \pm 165$ & J0536+0944 \\
GBS-VLA J053459.44+095312.0 & BL175GO & 2456733.58 & $05\, 34\, 59.441635$ & $0.000014$ & $+09\, 53\, 11.91540$ & $0.00023$ & $1797 \pm 115$ & J0536+0944 \\
GBS-VLA J053459.44+095312.0 & BL175ER & 2456950.00 & $05\, 34\, 59.441575$ & $0.000016$ & $+09\, 53\, 11.91674$ & $0.00034$ & $1492 \pm 159$ & J0536+0944 \\
GBS-VLA J053459.44+095312.0 & BL175FL & 2457096.60 & $05\, 34\, 59.441538$ & $0.000010$ & $+09\, 53\, 11.91680$ & $0.00034$ & $1057 \pm 115$ & J0536+0944 \\
GBS-VLA J053459.44+095312.0 & BL175FX & 2457294.01 & $05\, 34\, 59.441626$ & $0.000005$ & $+09\, 53\, 11.91970$ & $0.00011$ & $2210 \pm 200$ & J0536+0944 \\
GBS-VLA J053449.71+095153.8 & BL175GO & 2456733.58 & $05\, 34\, 49.707676$ & $0.000033$ & $+09\, 51\, 53.77293$ & $0.00075$ & $3156 \pm 384$ & J0536+0944 \\
\hline\hline
\end{tabular}
\tablefoot{Columns are (left to right): The source name as they appear in SIMBAD astronomical database  and/or previous radio surveys; in particular, GBS--VLA refers to sources identified in the Gould's Belt VLA Survey \citep{kounkel2014}, the VLBA code of the observation, the Julian date, the measured position in right ascension and declination with their respective uncertainties, the measured flux densities corrected by the primary beam attenuation, and the main phase calibrator used to observe the target. 
This table is published in its entirety at the CDS via anonymous ftp to
\texttt{cdsarc.u-strasbg.fr} (130.79.128.5) or via \url{https://cdsarc.cds.unistra.fr/viz-bin/cat/J/A+A}.
A portion is shown here for guidance regarding its form and content. }
\end{center}
\end{table*}

\subsection{Gaia DR3}

The \textit{Gaia} DR3  data \citep{gaia2023}, and data from previous catalogs, can be accessed from the 
Gaia archive\footnote{\url{https://gea.esac.esa.int/archive/}}. Using the tools
offered in the \textit{Gaia} archive, we searched for optical counterparts of our detected 
radio sources. For the cross-matching between radio and optical sources, we have
used a radius of $0\rlap{.}''5$. 

For the matched sources, we retrieved the \textit{Gaia} Renormalised Unit Weight Error (RUWE), which is a goodness-of-fit metric for the single-star astrometric solution. Values significantly larger than unity indicate potentially problematic solutions (e.g., unresolved binarity or calibration issues). In this work, RUWE is used solely as a quality indicator to identify such cases.

\section{Results} \label{sec:results}

In this Section, we present the astrometric results from our observations of the young stellar population with radio emission in the Orion complex. 
We summarize the detection statistics, describe the identification of background contaminants, provide astrometric solutions for 58 sources, and assess the impact of systematic position errors. The orbital solutions are presented and discussed in the companion \citetalias{dzib2025}.

\subsection{Detection statistics}\label{sec:statdet}

Due to the limited number of baselines and the high angular resolution of VLBI arrays, the resulting images often contain noise peaks exceeding 5$\sigma$, which can lead to false-positive detections \citep[see discussion in][]{forbrich2021}. To mitigate this, semi-blind VLBI surveys typically adopt a conservative detection threshold of ${\rm S/N} > 6$ \citep[e.g.,][]{herrera2017,forbrich2021}. In this work, we adopt a slightly lower, but still conservative, detection limit of 5.5$\sigma$. Sources detected in only one epoch with $5.5 < {\rm S/N} < 6.0$ are classified as candidate detections. In specific cases, we allow a relaxed threshold down to 5$\sigma$ when a source position is consistent, within the expected motion from proper motion and parallax, with that of a $\geq5.5\sigma$ detection from another epoch. In Table~\ref{tab:full}, we present the properties of all sources detected in Orion from both the GOBELINS and DYNAMO-VLBA projects. We note that, due to our more stringent detection criteria, some of the sources reported by \citetalias{kounkel2017} are not included in our list.

\begin{sidewaystable*}
\scriptsize
\setlength{\tabcolsep}{2.6pt}
\renewcommand{\arraystretch}{0.99}
\caption{VLBA and \textit{Gaia} parallaxes and proper motions.}\label{tab:par}
\centering
\begin{tabular}{ccccccccccccccccccc}
\hline\hline
      && & \multicolumn{6}{c}{VLBA}& &&& \multicolumn{4}{c}{Gaia}\\ \cline{4-9}\cline{13-16}
Name & SpT&N$_{\rm det}$& R.A. [2016.0] & Dec. [2016.0] & $\varpi$ & Distance & $\mu_{\alpha}^*$ & $\mu_{\delta}$ & $\Delta_{VG\alpha}$ & $\Delta_{VG\delta}$ & $\Delta_{VG,total}$ & $\varpi$ & $\mu_{\alpha}^*$  & $\mu_{\delta}$  & RUWE \\
  & & &(deg)&(deg)&(mas)&(pc)&(mas yr$^{-1}$)&(mas yr$^{-1}$) &(mas)&(mas)&(mas)&(mas)&(mas yr$^{-1}$)&(mas yr$^{-1}$) & \\
\hline
\rowcolor{lightgray}\multicolumn{15}{l}{NGC\,2068}\\                                                                    %
GBS\,J054643.62+000528.3  & ...&14 & 86.681758636(48) & 0.091202441(34) & $2.40 \pm 0.17$ & $417 \pm 29$ & $-0.75 \pm 0.08$ & $-0.86 \pm 0.06$ & $-0.02 \pm 0.01$ & $0.23 \pm 0.01$ & $0.23 \pm 0.01$ & $2.41 \pm 0.02$ & $-0.80 \pm 0.02$ & $-0.78 \pm 0.02$ & 1.1 \\ 
HD 290862  & B3/5&15 & 86.680773553(88) & 0.076676023(165) & $2.61 \pm 0.31$ & $383 \pm 46$ & $0.54 \pm 0.15$ & $-0.57 \pm 0.27$ & $-0.23 \pm 0.02$ & $0.28 \pm 0.04$ & $0.36 \pm 0.04$ & $2.36 \pm 0.04$ & $-0.02 \pm 0.05$ & $-0.55 \pm 0.04$ & 2.3 \\
$[$SSC75$]$ M\,78 11 &A0II& 16 & 86.688908026(40) & 0.044525815(27) & $2.48 \pm 0.13$ & $403 \pm 21$ & $0.32 \pm 0.07$ & $-0.98 \pm 0.05$ & $-0.23 \pm 0.01$ & $0.40 \pm 0.01$ & $0.46 \pm 0.01$ & $2.45 \pm 0.02$ & $0.17 \pm 0.02$ & $-0.99 \pm 0.01$ & 1.2 \\  
\rowcolor{lightgray}\multicolumn{15}{l}{NGC\,2024}\\
$[$BCB89$]$ IRS\,11& ...& 12 & 85.409371973(29) & --1.885901080(67) & $2.37 \pm 0.08$ & $423 \pm 14$ & $0.02 \pm 0.05$ & $-0.27 \pm 0.11$ & $0.31 \pm 0.01$ & $-0.15 \pm 0.02$ & $0.34 \pm 0.02$ & $2.46 \pm 0.19$ & $0.10 \pm 0.18$ & $-0.40 \pm 0.16$ & 0.9 \\ 
2MASS J05414134--01533260$^{a}$ &...&19/8 & 85.422302898(25) & --1.892455570(119) & $2.53 \pm 0.06$ & $396 \pm 9$ & $1.35 \pm 0.05$ & $-0.75 \pm 0.22$ & ...& ...& ...& ...& ...& ...& ... \\  
$[$BCB89$]$ IRS\,15& ...&17 & 85.407264414(39) & --1.897662194(44) & $2.50 \pm 0.13$ & $400 \pm 21$ & $-0.13 \pm 0.06$ & $-0.44 \pm 0.07$ & $2.08 \pm 0.01$ & $0.48\pm 0.01$ & $2.13 \pm 0.01$ & $2.51 \pm 0.07$ & $0.19 \pm 0.07$ & $-0.38 \pm 0.06$ & 1.1 \\    
CXOU J054145.8--015411 & ...&11 & 85.441259858(160) & --1.903104814(103) & $2.47 \pm 0.08$ & $405 \pm 13$ & $-0.38 \pm 0.03$ & $-0.50 \pm 0.11$ & ...& ...& ...& ...& ...& ...& ... \\
2MASS J05413786-0154323 & ...&11 & 85.407772400(48) & --1.908877676(52) & $2.38 \pm 0.14$ & $420 \pm 25$ & $-0.26 \pm 0.08$ & $-0.71 \pm 0.10$ & $0.18 \pm 0.01$ & $0.06 \pm 0.01$ & $0.19 \pm 0.01$ & $2.78 \pm 0.15$ & $-0.51 \pm 0.13$ & $-0.57 \pm 0.13$ & 0.9 \\ 
2MASS J05414138-0154445 & ...&5 & 85.422458006(167) & --1.912431516(186) & $2.39 \pm 0.44$ & $419 \pm 77$ & $1.00 \pm 0.54$ & $-2.17 \pm 0.61$ & ...& ...& ...& ...& ...& ...& ... \\ 
CXOU J054146.1-015622$^{a}$  & ...&20/17 & 85.442316034(25) & --1.939499847(150) & $2.56 \pm 0.09$ & $391 \pm 14$ & $0.14 \pm 0.01$ & $-0.83 \pm 0.20$ & ...& ...& ...& ...& ...& ...& ... \\
V* V621 Ori  & ...&7 & 85.469117322(39) & --2.064338636(21) & $2.47 \pm 0.12$ & $406 \pm 19$ & $0.62 \pm 0.13$ & $-1.13 \pm 0.07$ & $0.19 \pm 0.01$ & $-0.31 \pm 0.01$ & $0.37 \pm 0.01$ & $2.50 \pm 0.03$ & $0.65 \pm 0.02$ & $-1.31 \pm 0.02$ & 1.1 \\ 
\rowcolor{lightgray}\multicolumn{15}{l}{$\sigma$\,Orionis}\\
HD 294300$^{a}$  & G5e&5/11 & 84.902253176(217) & --2.704789011(98) & $2.48 \pm 0.07$ & $404 \pm 12$ & $1.95 \pm 0.07$ & $-0.48 \pm 0.21$ & $-1.57 \pm 0.02$ & $-0.31 \pm 0.02$ & $1.60 \pm 0.02$ & $3.00 \pm 0.07$ & $3.52 \pm 0.08$ & $-0.90 \pm 0.08$ & 5.0 \\
\rowcolor{lightgray}\multicolumn{15}{l}{ONC}\\
HD 37017$^{a}$& B2/3V&5 & 83.841118043(17) & --4.494166594(44) & $2.64 \pm 0.03$ & $379 \pm 4$ & $1.85 \pm 0.04$ & $1.21 \pm 0.10$ & $0.11 \pm 0.01$ & $0.23 \pm 0.01$ & $0.26 \pm 0.01$ & $2.78 \pm 0.07$ & $1.48 \pm 0.06$ & $1.55 \pm 0.04$ & 1.3 \\ 
V* V1727 Ori  & K6 & 4 & 83.801917776(40) & --4.740537823(145) & $2.24 \pm 0.14$ & $447 \pm 29$ & $2.51 \pm 0.22$ & $-1.35 \pm 0.78$ & $9.97 \pm 0.02$ & $0.57 \pm 0.04$ & $9.99 \pm 0.02$ & $2.39 \pm 0.28$ & $1.85 \pm 0.25$ & $-1.75 \pm 0.20$ & 17.9 \\              
V* V1699 Ori & M2.9 & 5 & 83.693405512(28) & --4.900561366(27) & $2.48 \pm 0.07$ & $404 \pm 11$ & $1.75 \pm 0.09$ & $-0.89 \pm 0.10$ & $0.02 \pm 0.01$ & $-0.35\pm 0.01$ & $0.35 \pm 0.01$ & $2.42 \pm 0.18$ & $1.62 \pm 0.16$ & $-0.80 \pm 0.13$ & 1.1 \\ 
V* V492 Ori & K8 & 4 & 83.844010568(259) & --5.133537615(17) & $2.59 \pm 0.57$ & $386 \pm 85$ & $1.59 \pm 0.35$ & $0.25 \pm 0.02$ & $-1.94 \pm 0.06$ & $-1.66 \pm 0.01$ & $2.55 \pm 0.03$ & $3.23 \pm 0.16$ & $0.90 \pm 0.14$ & $-0.37 \pm 0.12$ & 1.7 \\  
V* V1321 Ori & K0III& 5 & 83.767921688(38) & --5.136839750(72) & $2.48 \pm 0.09$ & $404 \pm 15$ & $0.05 \pm 0.13$ & $7.06 \pm 0.24$ & $-0.32 \pm 0.01$ & $-0.34 \pm 0.02$ & $0.47 \pm 0.01$ & $2.54 \pm 0.01$ & $0.26 \pm 0.01$ & $7.17 \pm 0.01$ & 1.1 \\ 
V* V363 Ori & K: & 5 & 83.919907841(13) & --5.150435655(84) & $2.63 \pm 0.07$ & $380 \pm 10$ & $0.44 \pm 0.05$ & $-0.86 \pm 0.28$ & $26.6 \pm 0.01$ & $-23.02 \pm 0.02$ & $35.18 \pm 0.01$ & $2.75 \pm 0.10$ & $2.06 \pm 0.09$ & $0.08 \pm 0.08$ & 8.9 \\ 
Brun 656$^{a}$  & G2III&17/7 & 83.838825715(44) & --5.203522220(103) & $2.48 \pm 0.06$ & $403 \pm 9$ & $1.73 \pm 0.04$ & $-0.36 \pm 0.13$ & $2.50 \pm 0.10$ & $-1.88 \pm 0.13$ & $3.13 \pm 0.11$ & $1.99 \pm 0.07$ & $2.53 \pm 0.07$ & $-4.28 \pm 0.06$ & 5.1 \\
V* NU Ori$^{a}$ &B0.5V& 28 & 83.880691065(124) & --5.267376172(182) & $2.44 \pm 0.10$ & $410 \pm 16$ & $1.46 \pm 0.07$ & $1.65 \pm 0.26$ & $-0.55 \pm 0.03$ & $0.35 \pm 0.04$ & $0.66 \pm 0.04$ & $2.41 \pm 0.06$ & $0.92 \pm 0.05$ & $1.72 \pm 0.04$ & 2.4 \\ 
V* V1230 Ori$^{b}$ &B1 & 9 & 83.836354438(27) & --5.362316314(70) & $2.51 \pm 0.04$ & $398 \pm 7$ & $-2.02 \pm 0.08$ & $4.20 \pm 0.13$ & $20.39 \pm 0.01$ & $29.40 \pm 0.02$ & $35.80 \pm 0.01$ & $2.46 \pm 0.03$ & $3.06 \pm 0.03$ & $-1.42 \pm 0.02$ & 2.0 \\ 
COUP 450$^{c}$ & K5V&16 & 83.799184034(15) & --5.363684562(21) & $2.58 \pm 0.04$ & $388 \pm 6$ & $1.57 \pm 0.04$ & $-1.41 \pm 0.05$ & ...& ...& ...& ...& ...& ...& ... \\ 
V* V1229 Ori & K0 & 16 & 83.826553477(16) & --5.377063296(22) & $2.57 \pm 0.06$ & $390 \pm 8$ & $2.30 \pm 0.03$ & $0.70 \pm 0.04$ & $-0.59 \pm 0.00$ & $-0.32 \pm 0.01$ & $0.68 \pm 0.01$ & $2.57 \pm 0.03$ & $2.15 \pm 0.03$ & $0.78 \pm 0.02$ & 1.0 \\   
ALMA\,J053514.50-052238.7$^{a}$  &...& 8 & 83.810423127(26) & -5.377416249(67) & $2.65 \pm 0.04$ & $378 \pm 5$ & $1.27 \pm 0.03$ & $-2.21 \pm 0.04$ & ...& ...& ...& ...& ...& ...& ... \\  
V* MT Ori$^{a}$  &K2& 15 & 83.824802068(48) & --5.379286952(72) & $2.66 \pm 0.03$ & $376 \pm 4$ & $3.57 \pm 0.05$ & $2.20 \pm 0.10$ & $-0.49 \pm 0.01$ & $0.78 \pm 0.01$ & $0.92 \pm 0.01$ & $2.48 \pm 0.04$ & $3.97 \pm 0.03$ & $2.74 \pm 0.03$ & 1.5 \\  
$\theta^1$ Ori E  & G2IV & 15 & 83.815723310(17) & --5.386075078(18) & $2.51 \pm 0.06$ & $399 \pm 10$ & $1.41 \pm 0.03$ & $1.06 \pm 0.04$ & $-0.42 \pm 0.00$ & $0.06 \pm 0.00$ & $0.42 \pm 0.00$ & $2.61 \pm 0.03$ & $1.34 \pm 0.02$ & $1.19 \pm 0.02$ & 1.5 \\ 
V* KM Ori & K5Ve&7 & 83.733225019(55) & --5.386955233(20) & $2.47 \pm 0.16$ & $404 \pm 27$ & $1.78 \pm 0.09$ & $0.04 \pm 0.04$ & $-0.45 \pm 0.01$ & $0.06 \pm 0.01$ & $0.46 \pm 0.01$ & $2.52 \pm 0.02$ & $1.45 \pm 0.02$ & $0.18 \pm 0.01$ & 1.2 \\  
$\theta^1$ Ori A$_2$ & ...& 16 & 83.815954882(13) & --5.387264298(25) & $2.53 \pm 0.04$ & $394 \pm 7$ & $4.81 \pm 0.02$ & $-2.48 \pm 0.05$ & $34.77 \pm 0.01$ & $179.23 \pm 0.01$ & $182.57 \pm 0.01$ & $2.64 \pm 0.07$ & $1.36 \pm 0.06$ & $0.25 \pm 0.05$ & 1.5 \\  
V* V1399 Ori$^{a}$&G8& 10 & 83.837705511(43) & --5.396945588(100) & $2.42 \pm 0.09$ & $413 \pm 15$ & $0.37 \pm 0.02$ & $0.46 \pm 0.14$ & $-0.39 \pm 0.07$ & $0.14 \pm 0.04$ & $0.41 \pm 0.05$ & $2.52 \pm 0.06$ & $0.44 \pm 0.06$ & $0.79 \pm 0.05$ & 3.2 \\  
Z05351603 &...& 5 & 83.816832315(62) & --5.398072705(15) & $2.72 \pm 0.21$ & $368 \pm 28$ & $2.26 \pm 0.19$ & $1.27 \pm 0.11$ & ...& ...& ...& ...& ...& ...& ... \\
Parenago 1540$^{a}$&K3V+K5V& 5 & 83.665665534(12) & --5.407112180(22) & $2.50 \pm 0.07$ & $400 \pm 11$ & $-3.92 \pm 0.08$ & $-1.12 \pm 0.06$ & $-1.43 \pm 0.01$ & $1.04 \pm 0.01$ & $1.77 \pm 0.01$ & $2.57 \pm 0.01$ & $-3.74 \pm 0.01$ & $-0.89 \pm 0.01$ & 1.0 \\ 
VLBA 13 &...& 4 &83.84532288(24) & --5.41600310(23) & $3.58 \pm 1.06$ &$279\pm83$ & $0.61 \pm 0.46$ & $4.12 \pm 0.44$ & ... & ... &... & ... & ...& ...&... \\
V* V1501 Ori& K4 - M1& 10 & 83.814812353(12) & --5.420590307(30) & $2.51 \pm 0.03$ & $398 \pm 5$ & $1.64 \pm 0.03$ & $1.59 \pm 0.07$ & $-0.21 \pm 0.01$ & $0.22 \pm 0.01$ & $0.30 \pm 0.01$ & $2.51 \pm 0.05$ & $1.53 \pm 0.05$ & $1.37 \pm 0.04$ & 1.3 \\ 
HD\,37150 & B3III/IV&5 & 84.062621199(29) & --5.647921522(23) & $2.56 \pm 0.08$ & $390 \pm 11$ & $1.30 \pm 0.10$ & $-0.57 \pm 0.10$ & $-0.45 \pm 0.01$ & $0.56 \pm 0.01$ & $0.72 \pm 0.01$ & $2.66 \pm 0.05$ & $1.21 \pm 0.05$ & $-0.15 \pm 0.04$ & 1.0 \\ 
\rowcolor{lightgray}\multicolumn{15}{l}{L1641}\\ 
TYC 5346-538-1 &B8.1& 6 & 85.640320325(27) & --8.120884216(71) & $2.31 \pm 0.09$ & $433 \pm 17$ & $0.67 \pm 0.10$ & $-0.31 \pm 0.25$ & $0.15 \pm 0.01$ & $-0.52 \pm 0.02$ & $0.54 \pm 0.01$ & $2.36 \pm 0.01$ & $0.58 \pm 0.01$ & $-0.19 \pm 0.01$ & 0.9 \\ 
2MASS\,J05420800--0812028 &M2.2& 6 & 85.533253837(28) & --8.200832577(107) & $2.26 \pm 0.09$ & $442 \pm 17$ & $0.13 \pm 0.10$ & $-0.88 \pm 0.37$ & $-1.01 \pm 0.03$ & $-0.34 \pm 0.04$ & $1.06 \pm 0.03$ & $2.23 \pm 0.63$ & $-0.33 \pm 0.56$ & $-1.45 \pm 0.53$ & 1.1 \\ 
\hline\hline
\end{tabular}
\vspace{-0.2cm}
\tablefoot{Columns are (left to right): Source name, R.A. and Dec. columns give the source coordinates at epoch J2016.0 from the VLBA astrometric solutions (after applying the calibrator-position corrections described in Sect.~3.4). $\Delta_{\rm VG\alpha}$ and $\Delta_{\rm VG\delta}$ are the positional differences between VLBA and \textit{Gaia} at J2016.0, expressed in mas and defined as VLBA $-$ Gaia; $\Delta_{\rm VG,total}=\sqrt{\Delta_{\rm VG\alpha}^{2}+\Delta_{\rm VG\delta}^{2}}$. \textit{Gaia} positions in the epoch J2016.0 are taken directly from the \textit{Gaia} DR3 catalog. 
RUWE is the \textit{Gaia} renormalised unit weight error. \\
$^{a}$ Astrometric parameters derived by including orbital parameters in the fitted model. Orbital parameters are presented and discussed in \citetalias{dzib2025}.\\
$^{b}$ Astrometric parameters derived by including acceleration terms in the fitted model. Resulted in (\acca,  \accd)=($-0.60\pm0.06$,\,$-1.80\pm0.11$) [mas\,yr$^{-2}$].\\
$^{c}$ Astrometric parameters derived by including acceleration terms in the fitted model. Resulted in (\acca,  \accd)=($-0.14\pm0.04$,\,$0.08\pm0.05$) [mas\,yr$^{-2}$].\\
}
\end{sidewaystable*}
By combining DYNAMO-VLBA observations with previous and new results from GOBELINS, we detect a total of 200 distinct radio sources across various regions of Orion. Of these, 77 sources (39\%) were detected in only one epoch, 44 (22\%) in two epochs, 21 (10\%) in three epochs, and 58 sources (29\%) in more than three epochs. Table~\ref{tab:full} lists each source along with the observing project, Julian date of observation, equatorial coordinates (right ascension and declination), statistical position errors derived from \texttt{jmfit}, and measured flux densities with their uncertainties. Multiple entries for a given source correspond to observations from different VLBA projects, all of which are indicated in the table.

Some sources are known to be YSOs. However, the nature of several radio sources remains uncertain. To investigate this, we analyze the spread of source positions across multiple epochs. We note that position differences are expected for sources at the distances of the Orion star-forming regions. First, the intrinsic proper motions, for example, in the case of the ONC, are on the order of a few mas per year \citep{dzib2017,dzib2021}. Additionally, there is the periodic apparent motion associated with the trigonometric parallax. For Orion sources (R.A. $\simeq$5-6 h), the dominant parallactic displacement has a peak-to-peak amplitude of $2\varpi\simeq$5\,mas and reaches its extrema close to the equinoxes, owing to the geometry of the parallax factors for this sky position. DYNAMO-VLBA, on the other hand, employed a different cadence for observations, with a minimum separation of two months; however, position changes could still be noticeable.

\subsection{Astrometric results}

The main focus of this paper is to determine accurate trigonometric parallaxes and proper motions for compact radio sources in the Orion complex. We obtained astrometric solutions for single radio sources associated with YSOs that are not known to belong to multiple systems, as well as for several radio sources that are members of binary or multiple systems.  Although the orbital analysis of these binaries will be presented in the companion \citetalias{dzib2025}, we include their measured parallaxes and proper motions here for completeness.

In total, 22 radio sources associated with YSOs and detected in four or more epochs were modeled with parallax and linear proper-motion fits. For five additional sources, the linear solutions yielded relatively large post-fit residuals showing clear periodic variations, suggesting binarity with orbital periods comparable to the VLBA time baseline. Figure~\ref{fig:NL} illustrates one such case, V*~V1399~Ori, previously not reported as a binary system, where our astrometric residuals reveal a periodic variation consistent with orbital motion. We repeat the fit to these sources and including orbital components, which resulted in substantially improved astrometric solutions.  A further 13 radio sources correspond to components of 9 known binary or multiple systems. Together, these 14 multiple systems are the subject of \citetalias{dzib2025}, where they are analyzed in detail.  The resulting astrometric parameters are listed in Table~\ref{tab:par}.  The measured trigonometric parallaxes range from 2.26 to 2.65\,mas, consistent with membership in the Orion complex.

\begin{table*}
\scriptsize
\setlength{\tabcolsep}{1.8pt}
\renewcommand{\arraystretch}{1.05}
\begin{minipage}{\linewidth}
\centering
\caption{VLBA proper motions of YSOs detected in two or three epochs by assuming a trigonometric parallax of $\varpi=2.50\pm0.35$\,mas.}\label{tab:pms}
\begin{tabular}{ccccccccccccccccccc}
\hline\hline
      &&& \multicolumn{4}{c}{VLBA}& &&& \multicolumn{3}{c}{Gaia}\\ \cline{4-7}\cline{11-13}
Name & SpT & N$_{\rm det}$& R.A. [2016.0] & Dec. [2016.0] & $\mu_{\alpha}^*$ & $\mu_{\delta}$ & $\Delta_{VG\alpha}$ & $\Delta_{VG\delta}$ & $\Delta_{VG,total}$ &  $\mu_{\alpha}^*$  & $\mu_{\delta}$  & RUWE \\
  & & &(deg)&(deg)&(mas yr$^{-1}$)&(mas yr$^{-1}$) &(mas)&(mas)&(mas)&(mas yr$^{-1}$)&(mas yr$^{-1}$) & \\
\hline
2MASS J05414412-0153472  &...& 3 & 85.43384506(08) & --1.89651549(08) & $0.3 \pm 0.2$ & $-0.0 \pm 0.2$ & ... & ... & ... & ... & ... &... \\ 
Brun 555 &K3& 2 & 83.81304162(44) & --4.74524606(30) & $18.3 \pm 1.2$ & $32.5 \pm 0.8$ & $18.24 \pm 0.10$ & $34.62 \pm 0.07$ & $39.13 \pm 0.08$ & $1.31 \pm 0.05$ & $0.81 \pm 0.04$ & 3.5 \\ 
HD 294264  &B2.5& 2 & 83.80560958(49) & --4.86248348(44) & $2.6 \pm 1.4$ & $-1.8 \pm 1.2$ & $0.78 \pm 0.12$ & $-0.68 \pm 0.10$ & $1.04 \pm 0.11$ & $1.42 \pm 0.04$ & $-1.44 \pm 0.03$ & 2.7 \\ 
2MASS J05345988-0455272 &...& 2 & 83.74955618(21) & --4.92422750(15) & $3.7 \pm 0.4$ & $1.7 \pm 0.3$ & ... & ... & ... & ... & ... &... \\ 
2MASS J05351956-0517032 &...& 2 & 83.83153194(41) & --5.28421615(65) & $-1.2 \pm 0.7$ & $-1.7 \pm 1.1$ & $12.46 \pm 0.11$ & $-10.53 \pm 0.16$ & $16.32 \pm 0.14$ & $2.27 \pm 0.99$ & $0.82 \pm 0.80$ & 1.2 \\
2MASS J05352349-0520016$^a$ & ...&2 & 83.84789462(13) & --5.33379943(13) & $1.8 \pm 0.2$ & $0.5 \pm 0.2$ & $-4.71 \pm 0.23$ & $3.84 \pm 0.20$ & $6.08 \pm 0.22$ & ... & ... &... \\  
JW 286 &K5& 3 & 83.77618035(06) & -5.36740706(03) & $0.7 \pm 0.1$ & $-0.6 \pm 0.1$ & $3.84 \pm 0.01$ & $4.55 \pm 0.01$ & $5.95 \pm 0.01$ & $0.70 \pm 0.09$ & $-0.56 \pm 0.07$ & 0.8 \\ 
ALMA J053514.6621-052211.277 &...& 2 & 83.81110509(17) & --5.36980058(14) & $-1.3 \pm 0.4$ & $-2.8 \pm 0.4$ & ... & ... & ... & ... & ... &... \\ 
ALMA J053514.8988-052225.416 &...& 3 & 83.81207696(08) & --5.37372460(04) & $-1.4 \pm 0.1$ & $-4.3 \pm 0.2$ & ... & ... & ... & ... & ... &... \\ 
ALMA J053514.5010-052238.674 &...& 3 & 83.81042269(31) & --5.37741760(28) & $1.6 \pm 0.4$ & $0.3 \pm 0.4$ & ... & ... & ... & ... & ... &... \\ 
COUP 530  &...& 3 & 83.80503553(10) & -5.38189009(08) & $0.3 \pm 0.2$ & $-2.3 \pm 0.2$ & ... & ... & ... & ... & ... &... \\  
COUP 625 & ...&3 & 83.80973079(05) & --5.38817293(03) & $3.3 \pm 0.1$ & $-0.8 \pm 0.1$ & ... & ... & ... & ... & ... &... \\ 
V* V1326 Ori &K8& 2 & 83.79070594(09) & --5.39080281(02) & $1.3 \pm 0.5$ & $0.3 \pm 0.3$ & $-0.34 \pm 0.02$ & $-0.03\pm0.01$ & $0.34\pm0.01$ & $1.20 \pm 0.03$ & $0.52 \pm 0.02$ & 1.2 \\ 
V* LQ Ori &K8& 2 & 83.79471978(09) & --5.39573739(02) & $-4.4 \pm 0.4$ & $-0.7 \pm 0.2$ & $-14.73 \pm 0.02$ & $27.97 \pm 0.01$ & $31.61 \pm 0.01$ & $2.00 \pm 0.05$ & $0.74 \pm 0.04$ & 1.7 \\ 
V* LQ Ori &K8& 2 & 83.79472640(22) & --5.39574694(10) & $-2.1 \pm 0.9$ & $6.0 \pm 0.4$ & $9.12 \pm 0.05$ & $-6.40 \pm 0.02$ & $11.14 \pm 0.04$ & $2.00 \pm 0.05$ & $0.74 \pm 0.04$ & 1.7 \\ 
2MASS J05350968--0523558 &...& 2 & 83.79031399(13) & --5.39886456(10) & $-0.3 \pm 0.1$ & $0.4 \pm 0.1$ & ... & ... & ... & ... & ... &... \\  
V* V1961 Ori &G9IV& 3 & 83.61390874(12) & --5.40619420(05) & $-7.1 \pm 0.3$ & $-1.0 \pm 0.1$ & $-1.02 \pm 0.03$ & $1.21 \pm 0.01$ & $1.58 \pm 0.03$ & $-7.29 \pm 0.02$ & $-0.90 \pm 0.01$ & 1.2 \\ 
V* KO Ori  & A7 &2 & 83.73534438(19) & --5.52671762(08) & $1.4 \pm 0.3$ & $-1.6 \pm 0.1$ & $-0.14 \pm 0.05$ & $-0.24 \pm 0.02$ & $0.28 \pm 0.02$ & $1.01 \pm 0.02$ & $-1.43 \pm 0.01$ & 1.1 \\ 
\hline\hline
\end{tabular}
\tablefoot{$^a$ Source with counterpart in \textit{Gaia} DR3 catalog, but without derived parallax and proper motions.}
\end{minipage}
\end{table*}

Following a similar approach to \citet{galli2018}, for each source, we compute the unique pairwise position (absolute) differences between detections in different epochs and then estimate the mean of these differences. In Fig.~\ref{fig:Sep} we plot the histograms of the estimated separations. We note that the histograms are strongly peaked at separations below 
$\sim1.5$\,mas, with a clear tail toward larger values corresponding to sources with measurable astrometric motion. We therefore adopt 1.5\,mas as an empirical threshold separating sources whose inter-epoch positional changes are consistent with the typical residual astrometric uncertainties from those exhibiting statistically significant motion. In this context, sources with position changes $<1.5$\,mas are likely background sources and will be discussed in Section~\ref{sec:background}. 
Sources with larger positional shifts are likely young stars in star-forming regions of Orion. 

An additional 18 sources are detected only two or three times and show significant positional displacements (see Sect.~\ref{sec:statdet}). For these, we determined barycentric positions at the reference epoch 2016.0 and fitted proper motions by fixing the parallax to $\pi=2.5\pm0.35$\,mas, adopted as a representative value within the distance range of the Orion subregions. These sources are listed in Table~\ref{tab:pms}.

\subsection{Background sources}\label{sec:background}

There are radio sources whose positions remain consistent across different observed epochs, with mean inter-epoch positional differences $\lesssim$1.5 mas(see Fig. 1), consistent with the expected level of systematic errors (Sect. 3.4), indicating that they are likely background Galactic or extragalactic sources.
Regarding GOBELINS results, \citet{kounkel2017} noticed the following cases\footnote{The source numbers quoted here follow the VLBA numbering scheme introduced by Kounkel et al. (2017) and can be directly cross-matched with the source names given in Table~\ref{tab:full}} VLBA 39, 40, 41, 42, 47, 48, 55, 56, 83, 84, 85, 88, 89, 90, 94, 95, 96, 146, 147. In addition to these sources, we also noted that the following radio sources have no indications of measurable parallaxes and proper motions: VLBA 1, 2, 15, 17, 21, 23, 24, 25, 29, 30, 31, 35, 36, 38, 43, 44, 49, 52, 54, 57, 65, 66, 70, 71, 72, 73, 74, 75, 76, 77, 78, 79, 87, 91, 92, 97, 98, 99, 100, 101, 102, 128, 139, 144, 151, 156, and 166. These sources are also listed in Table~\ref{tab:full} for completeness; however, we did not conduct further analysis on them.

The case of VLBA 2 is noteworthy. It is a background radio source, likely extragalactic, located $\sim9\rlap{.}'5$ from the center of the ONC and has a flux density of 5\,mJy, and a variability of $\sim10\%$. While its flux density is too low to serve as an MPC, it could be used for additional phase corrections in future VLBI observations of the ONC to improve phase calibration. Such a strategy has successfully been used in the past for other VLBA observations \citep[e.g.,][]{dzib2013cyg}.

\subsection{Systematic position errors}

In modeling the motion of the radio sources, we initially account for only the statistical position uncertainties, including the contribution from thermal noise, derived from {\tt JMFIT}. For weak sources, this thermal-noise component dominates the formal positional uncertainty. However, as noted in Section~\ref{sec:data}, these errors reflect image-fitting uncertainties alone and do not account for additional sources of error, such as phase transfer imperfections. Estimating the magnitude and origin of such systematic errors is inherently challenging.

To compensate for residual position errors introduced during phase transfer, we adopted a procedure in which systematic uncertainties were added in quadrature to the statistical errors until the reduced $\chi^2_{\rm red} = 1.0$ was achieved in the astrometric fits. This adjustment was performed independently for each source, depending on its flux density and number of detections. On average, the additional systematic errors required were $\sigma_{\alpha, \rm sys} = 0.27$\,mas and $\sigma_{\delta, \rm sys} = 0.39$\,mas. Because these are post-fit residuals conditioned on the best-fit astrometric model, their distribution is expected to be narrower than that of unconditioned Gaussian measurement errors, and a high fraction of points within $\pm1\sigma$ does not imply overestimated uncertainties.

\begin{figure}
    \centering
    \includegraphics[width=0.66\linewidth]{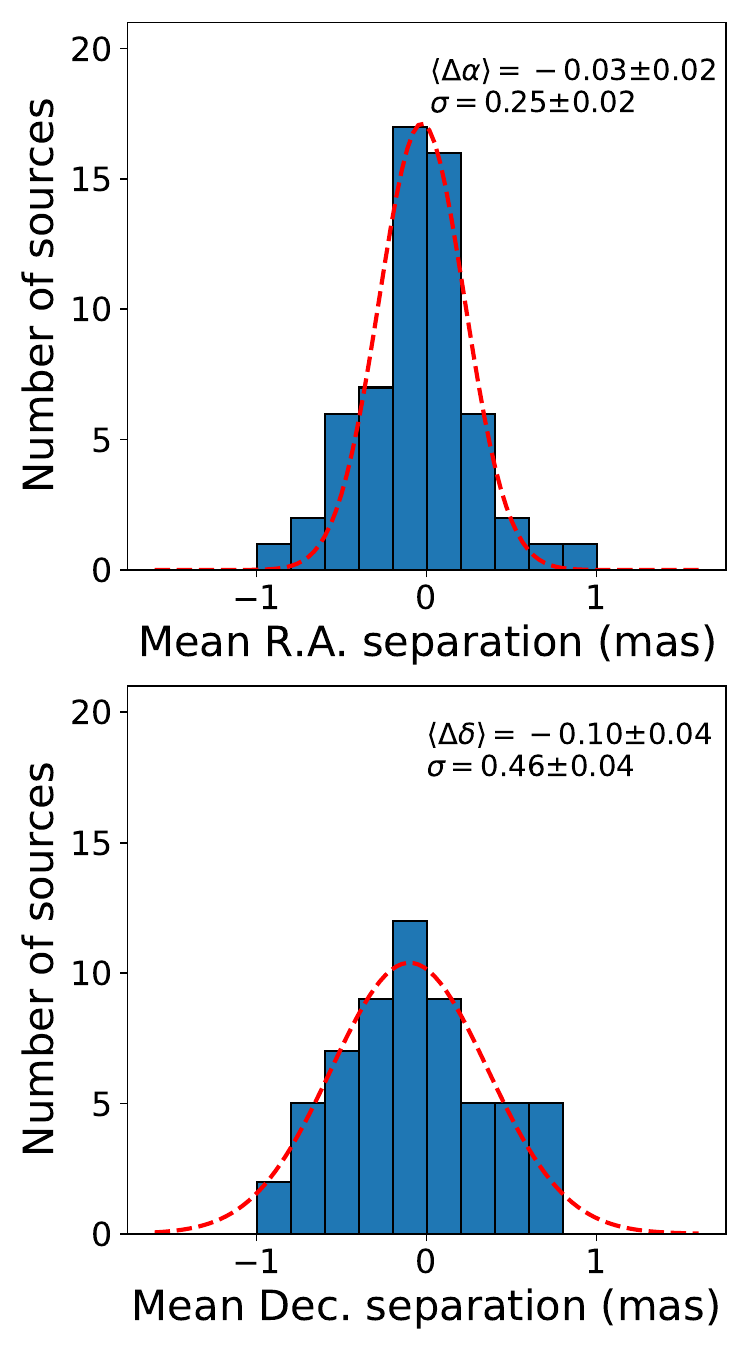}
    \caption{Histograms showing the pairwise separation of background sources through epochs.}
    \label{fig:Sepq}
\end{figure}

We further assessed these systematic errors using a subset of background radio sources identified in previous sections. Assuming these are predominantly extragalactic (and thus have negligible motion), any measured positional offsets between epochs can be interpreted as an empirical estimate of residual systematics. Figure~\ref{fig:Sepq} shows histograms of the measured position differences for these sources, with dispersions of 0.25\,mas in right ascension and 0.46\,mas in declination, values consistent with those added during our astrometric fits. This agreement supports the robustness of the systematic error model we adopted.

Additionally, we note that the positions of the Main Phase Calibrators (MPCs) used in this work were not corrected using the latest available reference frame solutions. Since target positions are directly tied to the MPC positions, this implies that target coordinates also inherit any systematic offsets. For consistency with future comparisons, we have opted to report uncorrected positions in Table~\ref{tab:full}, while the MPC positions and differences with current catalogs are listed in Table~\ref{tab:PhCal}.

We justify this choice by noting that radio calibrator catalogs are continuously refined with new VLBI observations. Providing uncorrected target positions allows for more consistent future comparisons, using whatever MPC positions are available at the time. Nevertheless, we emphasize that when comparing our astrometry with external datasets (e.g., Gaia), the MPC position offsets should be accounted for. Therefore, after performing the astrometric fits, we apply the corresponding current MPC correction (see Table~\ref{tab:PhCal}) to the final derived positions (at epoch J2016.0).

\section{Discussion}\label{sec:discussion}

Our VLBA observations of young stellar objects in Orion provide one of the most comprehensive astrometric datasets available for this region. The precision of these measurements enables us not only to trace the internal kinematics of individual stars and multiple systems, but also to derive robust dynamical masses and refine the distances to the main star-forming regions. In this section, we discuss the implications of our results in three contexts. First, we examine the orbital solutions of binary and multiple systems and compare the derived dynamical masses with predictions from stellar evolutionary models. Second, we utilize the parallaxes of individual sources to constrain the distances to different subregions of Orion, thereby placing them within the broader framework of the complex three-dimensional structure. Finally, we compare our VLBA results with those from \gaia\,DR3 to evaluate potential systematics in both datasets and to assess the consistency between radio and optical reference frames.

\begin{table*}[!h]
\setlength{\tabcolsep}{4pt}
\renewcommand{\arraystretch}{1.05}
    \centering
    \caption{Mean parallaxes and distances to Orion star-forming regions derived from VLBA (this work) and \gaia\,DR3 data.}
    \label{tab:OriD}
    \begin{tabular}{cccc|c|cccccc}
    \hline\hline
               &\multicolumn{3}{c|}{VLBA} &\citetalias{kounkel2017}& \multicolumn{6}{c}{\gaia DR3}\\
       Region  & \#   & $\bar{\varpi}$ & $\bar{d}$&$\bar{d}$&R.A.& Dec.&$\theta$ &\#   & $\bar{\varpi}$ & $\bar{d}$\\
       Name    & & (mas)       &(pc) & (pc) &($^{h}\,^{m}$) & ($^\circ\,'$) &($'$) & & (mas)   &     (pc)\\
       \hline
       NGC\,2068  & 3  & $2.47\pm0.10$ & $405\pm16$&$388\pm10$&  5 46.8 & +0 4.0 & 3&19 & $2.454\pm0.009$ & $407.5\pm1.5$\\
       NGC\,2024  & 8  & $2.48\pm0.03$ & $403\pm5$ &$423\pm15$&5 41.7 & --1 51.4 &15&160& $2.551\pm0.005$ & $392.0\pm0.7$ \\
$\sigma$\,Orionis & 1  & $2.48\pm0.07$ & $407\pm12$&$302\pm32^{\rm a}$&5 38.7 & --2 36.0& 10 &140& $2.463\pm0.004$ & $406.0\pm0.6$\\
       ONC        & 21 & $2.57\pm0.01$ & $388.5\pm1.7$ &$388\pm5$&5 36.3 & --5 23.4&15& 497 & $2.539\pm0.002$& $393.9\pm0.4$ \\
       L1641      &  2 & $2.29\pm0.06$ & $438\pm12$&$428\pm10$&5 42.6 & --8 12.0 &10&96&  $2.369\pm0.004$& $422.2\pm0.8$\\ 
       \hline
    \end{tabular}
    \tablefoot{$^{\rm a}$ \citetalias{kounkel2017} do not list this distance for the region. We added it here as it is related to the same source we analyzed. However, as noticed by those authors, the value is affected by the fact that the source was detected in three epochs and classified as a binary candidate. }
\end{table*}

\subsection{Distances to star-forming regions within Orion}

Precise distances are essential for characterizing stars, their environment,  and related phenomena. In star-forming regions, obtaining accurate distance estimates to all individual stars can be challenging due to high extinction, variability, and stellar crowding. Consequently, it is common practice to adopt a single representative distance per region, based on a subset of well-measured sources.

Using the parallaxes listed in Table~\ref{tab:full}, we computed weighted average parallaxes and distances for the five observed regions in Orion. The results are given in Table~\ref{tab:OriD}, together with previous VLBA estimates from \citetalias{kounkel2017}. Although the larger number of epochs in our data would normally lead to smaller uncertainties, the errors reported by \citetalias{kounkel2017} are often smaller. This is because their analysis does not account for residual systematic effects (e.g., phase gradients) and, in some cases, relies on only three detections, making their quoted uncertainties likely optimistic. For this reason, we do not interpret the comparison as a strict statistical consistency test based solely on the quoted uncertainties. Quantitatively, the distance differences (this work $-$ \citetalias{kounkel2017}; Table~\ref{tab:OriD}) are $+17$\,pc (NGC\,2068), $-20$\,pc (NGC\,2024), $+0.5$\,pc (ONC), and $+10$\,pc (L1641), corresponding to 0.9$\sigma$, 1.3$\sigma$, 0.1$\sigma$, and 0.6$\sigma$, respectively. For $\sigma$ Orionis, \citetalias{kounkel2017} do not provide a regional distance estimate, and the value listed in Table~\ref{tab:OriD} refers to a single problematic source; it is therefore not used for a region-level comparison.

We further compared our VLBA-based distances with independent estimates derived from \gaia\ DR3. We selected \textit{Gaia} sources with $2.222 \leq \varpi \leq 2.85$\,mas and RUWE $\leq 1.4$, and adopted region-dependent search radii encompassing the spatial extent of each complex. For L1641, owing to its elongated morphology and the small number of VLBA detections, we used the mean position of the two radio sources and a $10'$ search radius. The adopted parameters and resulting weighted averages are summarized in Table~\ref{tab:OriD}.

The VLBA and \textit{Gaia} parallaxes agree within $\lesssim2\sigma$ for most regions when the quoted uncertainties are combined in quadrature. The largest offsets occur for NGC\,2024 ($\sim2\sigma$ level) and for the ONC ($\sim3\sigma$). These differences should be interpreted cautiously, given the small number of regions and the presence of residual systematic effects in both datasets. In the case of NGC\,2024, local astrophysical effects such as internal substructure may also contribute \citep{vanterwisga2020}.

Finally, for each region we assessed whether the observed parallax dispersion indicates measurable line-of-sight depth. The expected dispersion $\sigma_{\rm exp}$ was estimated as the rms of the individual parallax uncertainties, and the intrinsic dispersion was defined as $\sigma_{\rm int}=\sqrt{\sigma_{\rm obs}^2-\sigma_{\rm exp}^2}$ when $\sigma_{\rm obs}>\sigma_{\rm exp}$, and zero otherwise. For both VLBA and \textit{Gaia} data, we find $\sigma_{\rm obs}\leq\sigma_{\rm exp}$ in all regions, yielding $\sigma_{\rm int}=0$ within the uncertainties. Thus, no statistically significant depth is detected, and the observed distance dispersions should be interpreted as upper limits on the physical extent along the line of sight.

\subsection{Comparison with \gaia}

The astrometric precision of VLBI observations provides a unique opportunity to assess and validate the accuracy of \gaia\,DR3 measurements. In this section, we directly compare the positions, parallaxes, and proper motions derived from our VLBI analysis with those reported by \gaia for sources common to both datasets. This comparison allows us to quantify systematic differences, examine possible offsets, and assess the consistency between the radio and optical reference frames. We first examine positional offsets, then extend the analysis to parallax and proper motion measurements.

\begin{figure}[!h]
    \centering
    \includegraphics[width=0.93\linewidth]{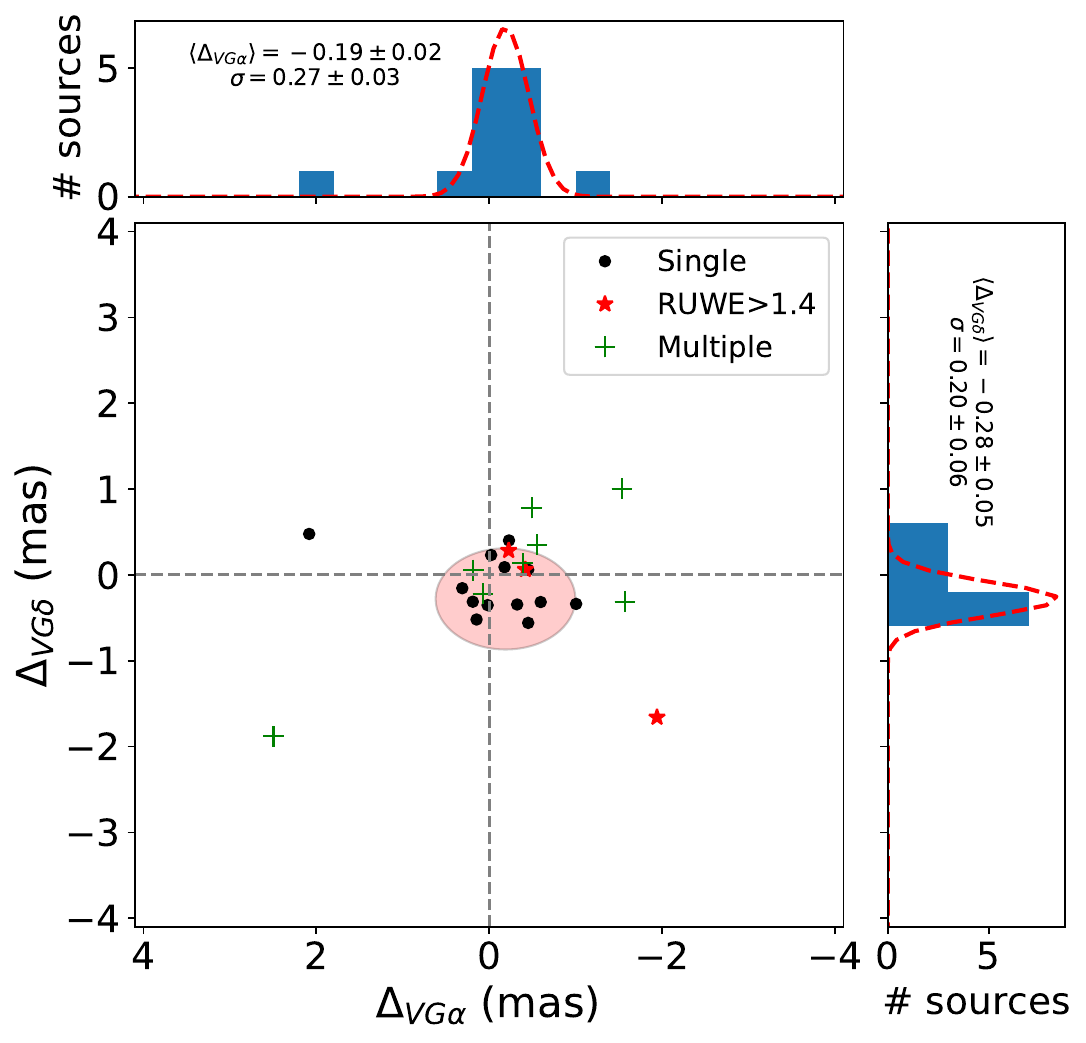}
    \caption{Distribution of the differences of \gaia and VLBA positions.
The central panel shows the offsets in right ascension and declination for three categories of sources: single stars (black circles), sources with RUWE\,$>1.4$ (red stars), and sources identified as multiple systems (green crosses). The red ellipse represents a 3$\sigma$ contour derived from a 2D Gaussian fit to the single-star and RUWE\,$\leq1.4$ distributions. The top and right panels display the corresponding histograms of offsets in right ascension and declination, respectively, for single stars, and RUWE\,$\leq1.4$. Dashed red lines show the best-fit Gaussian models, with the fitted mean and standard deviation indicated in each subplot.}
    \label{fig:VGpos}
\end{figure}

\begin{figure*}
    \centering
    \includegraphics[width=0.92\linewidth]{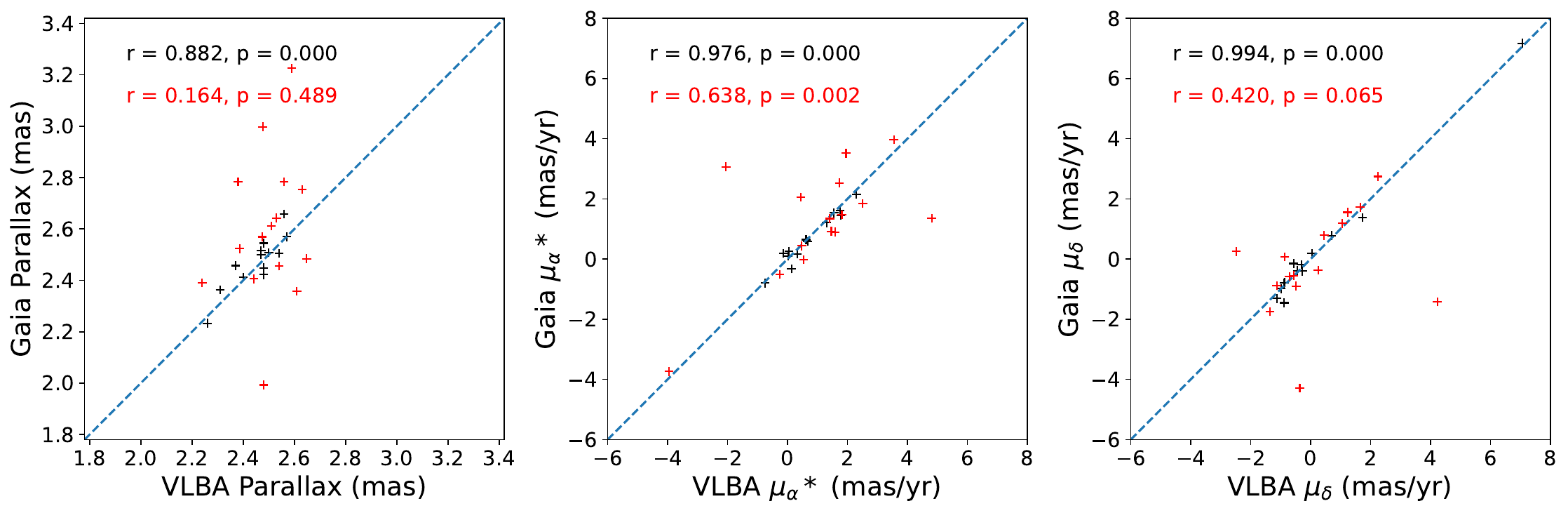}\\
    \includegraphics[width=0.92\linewidth]{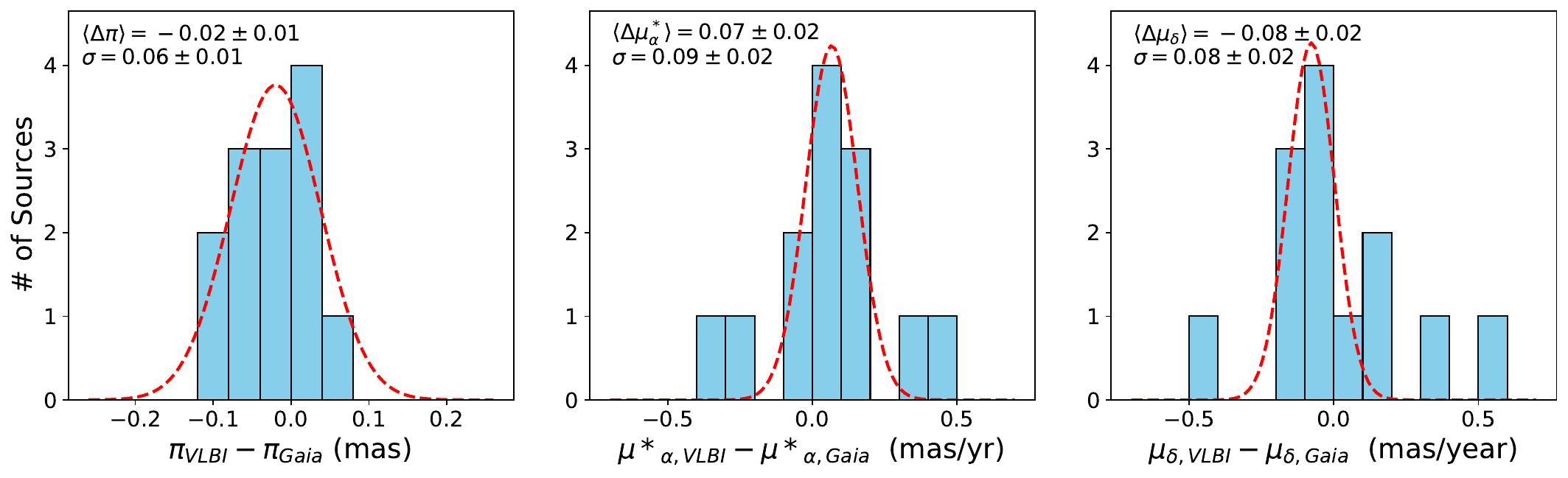}
    \caption{VLBA-\gaia parallax (left-panel) and proper motion ($\mu^*_\alpha$ in the center, and $\mu_\delta$ to the right) comparisons. {Top-panels:} Scatter plots of measured values. Red symbols indicate sources related to multiple systems, and black symbols are likely single sources. The dashed line is the equality line, which is included for eye guidance. Determined correlation coefficient (r) and its statistical significance are included in each case. { Lower panels:} Histograms showing the differences between VLBI and \textit{Gaia} EDR3 astrometric parameters for matched likely single sources. }
    \label{fig:VGpuu}
\end{figure*}

\subsubsection{Positions}

The astrometric precision achieved with VLBI is comparable to that of \gaia. For those radio sources with full astrometric solutions from our VLBI fitting (Table~\ref{tab:par}) and with counterparts in the \gaia archive, we conducted a direct comparison of the derived positions. We identified 28 such sources with entries in \gaia\,DR3. In Table~\ref{tab:par}, we report the VLBI--Gaia positional offsets in right ascension ($\Delta_{VG\alpha}$), declination ($\Delta_{VG\delta}$), and total separation ($\Delta_{VG,\rm total}$).

Previous studies have reported systematic offsets between VLBI and \gaia positions \citep[e.g.,][]{lindegren2020vlbi,dzib2021,lunz2023}. For example, \citet{dzib2021} found a systematic offset of $1.03\pm0.16$\,mas in declination between VLBA and \gaia\,DR2 positions in the ONC. In that work, the phase calibrator (J0541--0541) had a reported positional correction of 0.23\,mas which does not fully account for the observed discrepancy \citet{dzib2021}.

To re-examine this issue, we followed a similar approach to \citet{dzib2021} using our expanded and updated dataset. Figure~\ref{fig:VGpos} shows the distribution of the VLBI--Gaia position offsets. We classified the sources into three categories: (i) those known to be binary or multiple systems, for which astrometric solutions included orbital or acceleration terms (green points); (ii) sources with elevated RUWE values (RUWE$>1.4$), suggesting unresolved multiplicity or astrometric anomalies (red points); and (iii) likely single stars, with RUWE$\leq1.4$ and no additional orbital modeling (black points).

For the single-star sample, we constructed histograms of the position offsets in right ascension and declination and fitted Gaussian distributions. The resulting mean offsets are $-0.19$\,mas (R.A.) and $-0.28$\,mas (decl.), with standard deviations of 0.27 and 0.20\,mas, respectively, reflecting the overall scatter of the offset distributions. 

Individual sources can exhibit offsets at the milliarcsecond level (see Table~6), arising from a combination of measurement uncertainties, residual systematics, and source-specific effects. To assess whether a global systematic offset is present, the relevant quantity is the uncertainty on the mean offset, given by the standard error of the mean ($\sigma/\sqrt{N}$). For $N=12$ sources, this yields uncertainties of $\sim0.08$\,mas (R.A.) and $\sim0.06$\,mas (decl.). The measured mean offsets are therefore small compared to the overall scatter and indicate no evidence for a large global systematic positional offset between the VLBA and \gaia\ reference frames.

Based on the width of the offset distributions, we adopt a conservative threshold of 1.0\,mas (approximately $3\sigma$) to assess whether VLBA and \gaia\ sources are co-located; larger separations may indicate unresolved multiplicity or non-coincidence between the optical and radio emission. Finally, we note that both VLBI and \gaia\ astrometric calibrations have improved since the earlier analysis by \citet{dzib2021}, which likely contributes to the reduced level of systematic offsets in the present results.

\subsubsection{Parallax and proper motions}

A comparison between astrometric results from \gaia and VLBI is highly valuable for assessing consistency and identifying potential systematic effects. It is well established that \gaia parallaxes suffer from a zero-point offset, which leads to a systematic overestimation of distances. This offset was first identified in \gaia\,DR2 \citep{lindegren2018} and varies with sky position, G-band magnitude, and (BP--RP) color. For \gaia\,DR2, the zero-point offset was estimated to lie between $-29$ and $-80\,\mu$as \citep{lindegren2018}, whereas for \gaia\,EDR3/DR3 the global mean offset is closer to $-17\,\mu$as \citep{lindegren2021}.

Proper motion systematics in \gaia are more subtle. Although the global zero-point for proper motions is generally consistent with zero \citep{lindegren2021}, the reference frame of bright stars is known to exhibit rotation relative to the extragalactic reference frame, introducing systematic proper motion biases ranging from 10 to 80\,$\mu$as yr$^{-1}$ \citep{lindegren2020vlbi, cantat2021}.

VLBI has played a crucial role in validating the \gaia parallax zero point. For instance, \citet{Xu2019} compared parallaxes from \gaia\,DR2 with those from VLBI, finding a systematic offset of $-75 \pm 29\, \mu$as. More recently, \citet{ye2025} compared parallaxes from \gaia\,DR3 with those derived from HST, VLBI, and orbital solutions. Their best estimate of the parallax zero point came from the VLBI comparison, yielding an offset of $-14.8 \pm 10.6\,\mu$as, consistent with the expected DR3 offset.

In contrast, studies specifically quantifying systematic offsets in \gaia proper motions using VLBI are less common. While comparisons exist \citep[e.g.,][]{lindegren2021,dzib2021,cantat2021}, systematic frame offset analyses have primarily been carried out using a limited number of bright radio stars. Notably, \citet{lindegren2020vlbi} analyzed the spin of \gaia bright-star reference frame using VLBI positions, confirming a small but measurable rotation.

With this context in mind, we now compare VLBA parallaxes and proper motions with those from \gaia DR3 for sources with reliable matches. In the top panels of Fig.~\ref{fig:VGpuu}, we plot \gaia DR3 values against the corresponding VLBI results. Points are color-coded: red for sources likely in multiple systems (based on binary classification, RUWE\,$>1.4$, or position offsets $\Delta_{VG\, \rm total}>1.0$\,mas), and black for likely single stars.

We compute the Pearson correlation coefficient ($r$) for each case. For single stars, the correlation is strong ($r \geq 0.87$) and statistically significant, indicating excellent agreement between VLBI and \gaia. For sources in multiple systems, the correlation weakens, likely due to unmodeled orbital motion in the \gaia solution.

For the single-star subsample, we further analyze the differences in astrometric parameters between \textit{Gaia} and VLBI. The bottom panels of Fig.~\ref{fig:VGpuu} show histograms of these differences. We find a mean parallax offset of $-0.02 \pm 0.01$\,mas and a standard deviation of $0.06 \pm 0.01$\,mas, indicating consistency within uncertainties and no statistically significant offset. However, the proper motion differences are $+0.07 \pm 0.02$\,mas yr$^{-1}$ in right ascension and $-0.08 \pm 0.02$\,mas yr$^{-1}$ in declination. These offsets are statistically significant at the $3.5\sigma$ and $4\sigma$ levels, respectively. Although small, they may reflect residual rotation in the \gaia reference frame, as previously reported.

\subsection{Discussion on individual sources}

\paragraph{Brun\,555} is listed as the radio source with the largest proper motion in our sample. However, we point out that this proper motion is likely not real, but rather the result of detecting two distinct radio sources. This is supported by its positional shift with its \textit{Gaia} counterpart, which is of a similar magnitude to the measured proper motion. The proper motion of its \gaia counterpart is low and compared with the values of other stars in the region\citep{dzib2021}. Because its low number of detections (2), no further astrometric analysis is possible, and this star is not discussed among the binary stars in \citetalias{dzib2025}.

We report two radio sources associated with the star V*\,LQ\,Ori.  Each radio source was detected in two epochs, and, interestingly, none of them appears to coincide with the position of its \gaia counterpart. We note that \citet{forbrich2021} and \citet{dzib2021} previously reported a radio source related to this star and observed with the VLBA; their source [FMR2016] 37. Interestingly, the \gaia-radio position shift they reported is 56.8\,mas, which is different from those measured with the radio sources here reported. 

\citetalias{kounkel2017} reported two radio sources, VLBA\,81 and VLBA\,82, which we associate with the star HD\,288313B, the companion of the intermediate-mass star HD\,288313. The two sources were detected in the same observed epoch, with a separation of about $0\rlap{.}''1$, significantly smaller than  the known separation of $0\rlap{.}''7$ between the optical components of the system. Because both radio detections occurred in a single epoch, these authors did not discuss them further. Nevertheless, we notice that systems of this kind are of particular interest, as multi-epoch VLBA monitoring can yield full orbital solutions and dynamical masses (see, e.g., \citetalias{dzib2025}). 

[BCB89]\,IRS\,15 is known to be a YSO. The position changes of the associated radio source VLBA\,58, detected through 17 epochs, were nicely described by a linear motion. The associated \gaia source has RUWE=1.1, also consistent with single–star astrometry. However, several clues suggest that the system could, in fact, be multiple. First, in their first observed epoch, \citepalias{kounkel2017} reported three radio sources associated with this object, VLBA\,58 (S/N=15), VLBA\,59 (S/N=9.0), and VLBA\,60 (S/N=5.8). Although only VLBA\,58 was detected in subsequent epochs by either GOBELINS or DYNAMO–VLBA. Second, the position offset between the \gaia and the VLBA radio source is 2.13\,mas, exceeding the $1$\,mas co-location threshold established in our analysis above. Finally, the \gaia and VLBA proper motions in right ascension are inconsistent at a level $\sim3\sigma$. Even though [BCB89]\,IRS\,15 is likely a multiple system, its luminosity and mass appear dominated by a single component, which explains the nearly linear astrometric behavior observed in both datasets.

CXOU\,J054146.1-015622 is a visual radio binary where two components are clearly detected in several epochs. However, in the epoch BD215-J0, we detected only one radio source whose measured position falls in the middle of the expected positions of the two components detected in the other epochs. Thus, we could not clearly associate the radio source detected in this epoch with any of the radio sources known in this system. Consequently, we could not unambiguously associate this detection with either of the known components. It remains uncertain whether this additional emission originates from a third, unresolved companion or from transient magnetospheric activity between the stars, such as large-scale magnetic structures (e.g., helmet streamers; \citealt{massi2008}). In our catalog, we identified this radio source as VLBA\,61-62.

We associated four compact radio sources with the variable star V*\,V363\,Ori. Two of them, VLBA\,105 and VLBA\,158, were originally reported by \citetalias{kounkel2017}. VLBA\,105 was detected in a total of five epochs, including two new detections presented here. VLBA\,158 was detected in one epoch at a separation of $\sim17$\,mas from VLBA\,105. Two additional faint sources were detected in single epochs in our new observations, following the IAU naming convention: J053540.77481$-$050901.5226 (epoch BL175K0; S/N=6.3) and J053540.78209$-$050901.6597 (epoch BD215F4; S/N=5.8). None of these sources is coincident with the \gaia~DR3 position of V*\,V363\,Ori. The \gaia astrometric solution for this star has a high RUWE value (8.9), suggesting that its motion deviates from a simple linear model and may be affected by multiplicity. Given the relatively low S/N of several detections, we cannot exclude the possibility that some of the faintest sources are spurious. Nonetheless, the presence of multiple, spatially clustered detections around V*\,V363\,Ori makes this system an interesting candidate for future high-sensitivity monitoring to confirm its multiplicity.

The X-ray source COUP\,625 was associated with three distinct compact radio sources; VLBA\,109 (detected in three epochs), VLBA\,142 (detected in one epoch), and DYNAMO-VLBA\,J053514.3318--052317315 (detected in one epoch). Sources VLBA\,109 and VLBA\,142 can be associated to radio sources [FRM2016]\,177-1 and [FRM2016]\,177-2 reported by \citet{forbrich2021} and \citet{dzib2021}, also observed with the VLBA. 

We report the first detection of a mas-scale radio source associated with $\theta^1$\,Ori\,C. The radio source was detected at a S/N=6.7 in a single epoch. We notice that its position is offset by about 14\,mas to the east from the optical position of the massive star.  $\theta^1$\,Ori\,C  is known to have a companion with an eccentric orbit, with an orbital period of $11.26\pm0.05$\,years, and semi-major axis of $44\pm3$\,mas \citep{kraus2009}. Given the orbital configuration of this system \citep[see Fig.\,8 in][]{kraus2009} it is likely that the radio source is associated with this companion. However, the estimated mass of the companion, $\sim11$,\msun\ \citep{balega2014}, places it well above the mass range of magnetically active stars typically responsible for nonthermal radio emission. Additional multi-epoch VLBA observations will be required to confirm the nature of the detected source.

The radio source VLBA\,13, was detected in four epochs. A linear proper-motion fit does not adequately reproduce the positional changes, suggesting possible multiplicity or orbital motion. The nearest bright optical counterpart in the \gaia\,DR3 catalog is associated with the massive star $\theta^2$\,Ori\,A,  located at an angular separation of $\sim0\rlap{.}''41$ from the radio source; thus, it is not listed in Table~\ref{tab:par}. The position angle of the radio source relative to the massive star is $296^\circ$. \citet{preibisch1999} reported a 3--7\msun companion to $\theta^2$\,Ori\,A located at a separation of $383\pm10$\,mas with a position angle of $291\rlap{.}^\circ1\pm1\rlap{.}^\circ5$. Because of the similar configuration, we associate the radio source with this companion.

\section{Conclusions} \label{sec:conclusions}

This work presents new VLBA observations of young stellar objects in the Orion complex as part of the DYNAMO-VLBA project. 
The multi-epoch astrometry yields accurate parallaxes and proper motions for radio-emitting YSOs across several subregions of Orion, extending the reach of previous VLBI campaigns such as GOBELINS, offering improved astrometric solutions tied to the ICRF. The derived parallaxes confirm that Orion spans a depth of about forty parsecs along the line of sight, with the ONC being in the front edge of the complex at $\sim389\pm2$\,pc.

A direct comparison of VLBI astrometry with Gaia\,DR3 results for the subset of sources with reliable optical counterparts showed that the positional offsets between VLBI and \textit{Gaia} are small, with no significant systematic shift detected for the likely single-star sample. Parallax measurements were also consistent, showing a negligible mean offset of $-0.02\pm0.01$\,mas. These findings reinforce the excellent agreement between both techniques and validate the reported \textit{Gaia} zero-point correction for its DR3 catalog. However, we did find statistically significant differences in proper motions: $+0.07 \pm 0.02$\,mas\,yr$^{-1}$ in right ascension and $-0.08 \pm 0.02$\,mas\,yr$^{-1}$ in declination. These results may reflect small residual systematics in the \textit{Gaia} reference frame or physical effects related to unresolved binary motion in either dataset. 

The DYNAMO–VLBA dataset thus provides the most accurate radio-based distances yet obtained for Orion, which can be used as a base for future comparison with the coming \gaia catalogs.  By combining the present results with the forthcoming orbital analyses of multiple systems (\citetalias{dzib2025}), the DYNAMO project offers a comprehensive view of the structure and dynamics of young stars in one of the nearest and most important star-forming regions.

\section*{Data availability}
Table \ref{tab:full} is fully available in electronic form at the CDS via anonymous ftp to cdsarc.u-strasbg.fr (130.79.128.5) or via http://cdsweb.u-strasbg.fr/cgi-bin/qcat?J/A+A/.

\begin{acknowledgements}
S.A.D. acknowledges the M2FINDERS project from the European Research
Council (ERC) under the European Union's Horizon 2020 research and innovation programme
(grant No 101018682).
L.L. acknowledges the support of DGAPA-PAPIIT grant IN108324 and SECIHTI grant CBF-2025-I-109.
G.N.O.L. acknowledges the financial support provided by Secretaría de Ciencia, Humanidades, Tecnología e Innovación (Secihti) through grant CBF-2025-I-201.
P.A.B.G. acknowledges financial support from the São Paulo Research Foundation (FAPESP, grant: 2020/12518-8) and Conselho Nacional de Desenvolvimento Científico e Tecnológico (CNPq, grant: 303659/2024-6).
The National Radio Astronomy Observatory is a facility of the National Science Foundation operated under cooperative agreement by Associated Universities, Inc.
\end{acknowledgements}

\bibliographystyle{aa_url} 
\bibliography{references.bib}

\appendix

\section{Current calibrator positions}

The positions of calibrators are being refined over time.
The most recent values are listed in the Radio Fundamental 
Catalog (RFC) described by \citet{petrov2025}. Its most 
recent version is the 2025B and is available through
\url{https://astrogeo.org/rfc/}. In Table~\ref{tab:NPhCal}
we list the RFC\,2025B positions of the calibrators used 
in this work.

\begin{table}[h]
\small
\begin{center}
\renewcommand{\arraystretch}{1.0}
\caption{Calibrators and positions from the RFC 2025B.}
\begin{tabular}{c|cccccccccccccc}\hline\hline
Name&  R.A.&Dec.\\ 
\hline
J0539--0514& \rahms{05}{39}{59}{937156} &\decdms{-05}{14}{41}{30073}\\
J0529--0519& \rahms{05}{29}{53}{533504} &\decdms{-05}{19}{41}{61736}\\
J0541--0541& \rahms{05}{41}{38}{083374} &\decdms{-05}{41}{49}{42862}\\
J0532--0307& \rahms{05}{32}{07}{519326} &\decdms{-03}{07}{07}{03706}\\
J0558--0055& \rahms{05}{58}{44}{391588} &\decdms{-00}{55}{06}{92853} \\
J0600--0005& \rahms{06}{00}{03}{503390} &\decdms{-00}{05}{59}{03431} \\
J0552+0313&  \rahms{05}{52}{50}{101448} &\decdms{+03}{13}{27}{24421} \\
J0542--0913&  \rahms{05}{42}{55}{877478} &\decdms{-09}{13}{31}{00591} \\
\hline\hline
\end{tabular}\label{tab:NPhCal}
\end{center}
\end{table}

\end{document}